\documentclass[11pt,a4paper]{article}

\newlength{\dinwidth}
\newlength{\dinmargin}
\usepackage[T1]{fontenc}
\usepackage[latin2]{inputenc}
\usepackage{graphicx}
\usepackage[normalem]{ulem}

\usepackage{fancyhdr}
\usepackage{calc}
\usepackage{epsfig}
\usepackage{amssymb}
\usepackage{amsmath}
\usepackage{indentfirst}

\def\lc{l}
\def\sp{S}
\def\s0{\sigma_0}
\def\alphas{\alpha_{\mathrm{s}}}
\def\alphaem{\alpha_{\mathrm{em}}}
\def\Hi{d\sigma ^{(qg^*)}}
\def\tHi{d\tilde\sigma ^{(qg^*)}}
\def\qval{q_{\mathrm{val}}}
\def\qsea{q_{\mathrm{sea}}}
\def\bqsea{\bar{q}_{\mathrm{sea}}}
\def\alamt{A_{\mathrm{LT}}}
\def\fww{{\cal F}_{\mathrm{WW}}}
\def\fjh{{\cal F}_{\mathrm{JH}}}
\def\fbfkl{{\cal F}_{\mathrm{BFKL}}}
\def\fgauss{{\cal F}_{\mathrm{G}}}
\def\be{\begin{equation}}
\def\ee{\end{equation}}
\renewcommand{\vec}[1]{\mbox{\boldmath $#1$}}

\begin{document}

\titlepage
%
\begin{center}
\vspace*{2cm}
{\Large \bf 
Lam--Tung relation breaking in $Z^0$ hadroproduction as a probe \\[2mm]
of parton transverse momentum} \\
\vspace*{7mm}
Leszek Motyka$^{1,*}$, Mariusz Sadzikowski$^{1,\dag}$ and Tomasz Stebel$^{1,2,\ddag}$
\\
\vspace*{7mm}
$^1${\it 
Institute of Physics, Jagiellonian University\\
S.\ \L{}ojasiewicza 11,  30-348 Krak\'{o}w, Poland
}
\\[0.5em]
$^2${\it Institute of Nuclear Physics PAN\\
Radzikowskiego 152, 31-342 Krak\'ow, Poland}
\\[1em]
{\it $^*$leszek.motyka@uj.edu.pl \\
$^\dag$mariusz.sadzikowski@uj.edu.pl \\
$^\ddag$tomasz.stebel@uj.edu.pl}\\[5mm]
 
\today

\end{center}

\vspace*{2ex}

\begin{abstract}
The Lam--Tung relation breaking coefficient $\alamt = A_0 - A_2$  in the Drell--Yan dilepton angular distributions in the $Z^0$ boson mass region at the LHC is analyzed in the $k_T$-factorization approach. This observable was recently measured with high precision by ATLAS collaboration. Within the $k_T$-factorization approach we perform an approximate ${\cal O}(\alphaem\alpha_s^2)$ calculation of the off-shell parton hard matrix elements in which we include the leading tree level contributions of valence quarks and off-shell gluons: the $\qval g^* \to qZ^0$ channel and the $g^*g^* \to q\bar q Z^0$ channel. The resulting $\alamt$ exhibits high sensitivity to the gluon transverse momentum distribution (TMD). Several gluon TMDs are probed derived from the CCFM and BFKL evolution equations, and given by QCD-inspired phenomenological parameterizations. The ATLAS data favor a simple ``Weizs\"{a}cker--Williams'' (WW) hard gluon TMD with the asymptotic behavior of one-gluon exchange at large gluon transverse momenta and moderate~$x$. 
It is verified that the proposed approach with the WW~gluon TMD describes well also the $A_0$ and $A_2$ angular coefficients at the $Z^0$ peak, as well as the Drell--Yan dilepton mass distribution at lower masses. We conclude that inclusion of gluon transverse momentum effects improves description of the angular distributions of Drell--Yan dileptons and that the Drell--Yan scattering provides an excellent probe of the parton TMDs. 
\end{abstract}

\newpage

\section{Introduction and conclusions}
\label{Sec:1}

The Drell--Yan process \cite{DrellYan} is an excellent probe of the proton structure in proton-proton or proton--anti-proton collisions. In this process a lepton--antilepton pair is produced by an intermediate neutral electroweak boson: virtual $\gamma$ or by a quasi-real or virtual heavy boson $Z^0$. The lepton--antilepton pair distributions from this process can be well measured over a wide range of kinematical parameters providing data on the Drell--Yan structure functions $W_{\tau}$ that depend on the pair invariant mass, total transverse momentum and rapidity, see e.g.\ \cite{LT12}. Four independent structure functions $W_{\tau}$ parametrize the $\gamma^*$ or the parity conserving $Z^0$ contribution, and five more structure functions are necessary to describe the odd-parity $Z^0$ contributions. Various Drell--Yan observables were measured recently at the LHC, for example the Drell--Yan mass distribution \cite{DYgamma_atlas,CMS:2014jea}, the transverse momentum dependence of the $Z^0$ boson \cite{Aad:2014xaa,Khachatryan:2015oaa}, and the coefficients of the lepton angular distributions at the $Z^0$ peak \cite{CMSLT,ATLASZ0}. In particular the ATLAS collaboration measured with excellent precision all the Drell--Yan structure functions at the resonant $Z^0$ production peak \cite{ATLASZ0}, that is for $M_{l^+l^-} \simeq M_Z$, as functions of the lepton pair transverse momentum, $q_T$. Both the total cross-section and most of the measured structure functions were found consistent with the theoretical predictions of perturbative QCD at the next-to-next-to-leading order (NNLO) \cite{ATLASZ0,DYNNLO,Karlberg:2014qua}. A striking exception from this agreement was found, however, in the Lam--Tung combination of the structure functions proportional to a difference of the lepton angular distribution coefficients, $\alamt = A_0 - A_2$ (for the explicit definition see Sec.\ \ref{Sec:3a})  where the experimental measurement of $\alamt$ reaches about 0.15 for $q_T > M_Z$, and the NNLO QCD prediction provides about half of this value in this region of $q_T$. This discrepancy is clearly visible for about 10~experimental data points of $\alamt$ with typical experimental errors of about $0.01\, - \, 0.02$.

The Lam--Tung combination of Drell--Yan structure functions is particularly interesting probe of subtle QCD effects. It follows from the fact that it vanishes at the leading twist up to the NLO in the collinear approximation. Thus, the Lam--Tung relation breaking may occur through higher twist effects or by QCD effects at the NNLO and beyond. For this reason $\alamt$ has been considered to be a promising probe of higher twist effects in the Drell--Yan process at small lepton pair masses, and at very high hadron collision energies \cite{Berger:1979du,Brandenburg:1994wf,Eskola:1994py,MSS,Brzeminski:2016lwh}. The Lam--Tung relation breaking may occur however, also at the leading twist as a result of the parton transverse momenta \cite{Boer:1999mm,Boer:2006eq,Berger:2007jw,Peng:2015spa}. In particular it was demonstrated~\cite{Peng:2015spa} that the Lam--Tung relation breaks down when both the quark and antiquark carry non-zero transverse momenta at the quark--antiquark--electroweak boson vertex. In the collinear framework such transverse momenta may be generated at higher orders, when additional emissions occur in the hard matrix element. In the $k_T$-factorization framework \cite{GLR,BFKL}, however, already the incoming partons have non-zero transverse momenta. Hence the Lam--Tung breaking coefficient $\alamt$ is sensitive to the details of the transverse momentum distribution of the incoming partons and may be used to probe parton Transverse Momentum Distributions (TMDs). On the other hand, the angular distributions of dileptons in heavy electroweak boson hadroproductions may be used to probe anomalous couplings of the bosons to quarks \cite{RWW}, and a good control of the QCD effects in these observables is necessary to enhance the sensitivity of the probes.

The parton TMDs --- in particular the gluon TMD --- parameterize important properties of the proton structure and allow to test and improve the QCD evolution equations with transverse momentum dependence, e.g.\ the Balisky--Fadin--Kuraev--Lipatov (BFKL) \cite{BFKL,bfklrev} or Catani--Ciafaloni--Fiorani--Marchesini (CCFM) \cite{CCFM} evolution equations. The procedure of calculating hard matrix elements for the partons with non-zero~$k_T$ was studied in detail in classical papers by Catani, Ciafaloni and Hautmann \cite{CaCiaHaut}.
Since then, the formalism based on $k_T$-factorization (or the high energy factorization) was successfully applied to numerous processes in high energy hadron scattering.

In the standard form and notation, the gluon TMD ${\cal F}(x,k_T ^2,\mu_F)$, depends on the gluon $x$ and $k_T$, and on the factorization scale $\mu_F$. The parton TMDs provide valuable insight into proton structure and properties of QCD --- for a recent review see \cite{TMDrev}. Their accurate determination is also important as the $k_T$-factorization formalism when applicable, may represent the scattering process kinematics more accurately than collinear QCD at given order of perturbation theory. Thus with better control of parton TMDs a more precise description of hadron scattering should be possible. Currently there exist many parameterizations of the gluon TMD with significantly different properties and it is important to find and probe observables sensitive to parton TMDs and to constrain the distributions \cite{TMDrev}. It should be stressed that although the $k_T$-factorization (or high energy factorization) approach was initially proposed for small~$x$ physics, the concepts of parton TMDs and the hard matrix elements with partons that carry non-zero transverse momentum may be also used beyond the small~$x$ limit, see Sec.\ \ref{Sec:3c} for a more detailed discussion and the references. 

The Drell--Yan process was analyzed within the $k_T$-factorization framework in three main approaches. Hence, the forward Drell--Yan cross-sections may be described in terms of the color--dipole formulation \cite{QCDdipole} of the $k_T$-factorization framework \cite{Brodsky,Kopeliovich,GJ,GolecBiernat:2010de,Ducati,SchSz,MSS}. In this approach a quasi-Compton emission of the electroweak boson from a fast (collinear) quark scattering by an off-shell gluon exchange takes place. In the second approach adopted e.g.\ in Refs.\ \cite{Watt:2003vf,HHJ,Dooling:2014kia,Nefedov:2012cq,Baranov:2014ewa}, an off-shell quark and antiquark fusion into the electroweak boson is considered. In the former the gluon TMD enters, while in the latter the quark and antiquark TMDs are used. The $k_T$-factorization framework is quite successful in both the approaches in describing Drell--Yan observables integrated over the lepton angles, but it does not give satisfactory description of all the Drell--Yan structure functions. In the third approach the contribution to Drell--Yan process coming from two initial gluons with non-zero $k_T$, hence the $g^* g^*$ partonic channel contribution. This partonic channel enters in QCD at the order ${\cal O}(\alpha_s^2)$.  Inclusion of the $g^* g^*$ partonic channel in the electroweak boson production was performed first for the prompt photon hadroproduction \cite{BLZprompt}, see also \cite{promptRV} for a recent study, then the formalism was applied to the Drell--Yan process and the heavy electroweak boson hadroproduction \cite{DeakSchwen,BLZdy}. For the Drell--Yan scattering, however, the $g^*g^*$ contributions were considered only for the total cross section and not for the structure functions.


In conclusion, in this paper we analyze sensitivity of $\alamt$ to the shape of quark and gluon TMDs within the $k_T$-factorization framework. We take into account both the lowest order contribution to the process, given by a simple fusion of quark and antiquark that both carry non-zero transverse momentum $k_T$, and channels that enter at higher orders of QCD (assuming the fixed order perturbative expansion). Namely these are the already known quark--gluon channel, and the gluon--gluon channel computed in this paper for the DY structure functions. The calculations are performed in the high energy limit in which the parton evolution is driven by the gluon evolution. We consider gluon TMDs emerging from solutions of the $k_T$-dependent evolution equations, BFKL and CCFM, and coming from QCD inspired models.  
It turns out that none of the used existing quark and gluon TMDs gives satisfactory description of $\alamt$ at large $Z^0$ transverse momentum. However, one should note that for such kinematics the TMDs are probed at relatively high gluon $x\sim 0.1$, where existing parameterizations are practically unconstrained. On the other hand we show that the application of the used formalism is well justified in this kinematic region. We therefore introduce new simple QCD inspired model of the gluon TMD which provides a good estimate of the Lam--Tung relation breaking at large $Z^0$ transverse momentum, that is for $q_T > M_Z$ and a reasonably good estimate for smaller transverse momenta. This model is based on a simple concept of the gluon TMD at moderately small gluon $x$ and at large gluon $k_T$, to be driven by the Weizs\"{a}cker--Williams gluon emission from valence quarks, resulting in $\sim 1 /k_T ^2$ behavior of the gluon TMD ${\cal F}(x,k_T ^2,\mu_F)$ at large $k_T$. We verify that this ``Weizs\"acker--Williams'' model of the gluon TMD provides not only a reasonable description of $\alamt$ at the $Z^0$ peak, but also a good description of the Drell--Yan pair mass $M$ distribution shape, probed by recent ATLAS measurements in the range of 15~GeV~$< M <$~55~GeV. Furthermore we demonstrate that $q_T$-dependence of $\alamt$ at the $Z^0$ peak is highly sensitive to the shape of the gluon TMD at large $k_T$ and hence this observable may be used to constrain gluon TMD with competitive precision.

\section{Kinematics and notation}
\label{Sec:2}

\begin{figure}
\centering
\includegraphics[width=.45\textwidth]{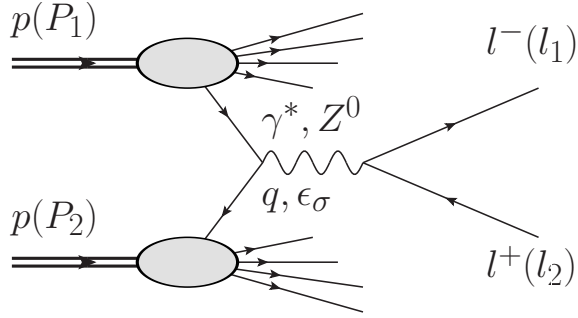} 
\caption{\it  The kinematics of the Drell--Yan process in $pp$ collisions. The diagrams in this paper are drawn with JaxoDraw \cite{JAXO}.
\label{Fig:kinem}}
\end{figure}

We consider the high energy proton--proton collision with a lepton--antilepton pair, $\lc^+ \lc^- = e^+ e^-$ or $\mu^+ \mu^-$, in the final state, $p(P_1)p(P_2) \to \lc^+ \lc^- X$,  and the leptons four-momenta $l^+$ and $l^-$ are measured, see Fig.\ \ref{Fig:kinem}. The lepton four-momenta are denoted by $l_1$ and $l_2$ for $l^-$ and $l^+$ correspondingly. At the leading order in the electroweak coupling constants this process is mediated by a virtual photon $\gamma^* (q)$  or by real or virtual $Z^0(q)$ boson, with the four-momentum $q = l_1 + l_2$, and $q^2 = M^2 > 0$ is the lepton pair invariant mass squared. 
The four-momenta of proton projectiles are $P_1$ and $P_2$, and they are near light-like --- in the center of mass system (c.m.s.) of the $pp$ pair and in the standard Minkovskian coordinates, one has $P_1 \simeq (\sqrt{\sp}/2,0,0,-\sqrt{\sp}/2)$, $P_2 \simeq (\sqrt{\sp}/2,0,0,\sqrt{\sp}/2)$, where the invariant collision energy squared $\sp=(P_1+P_2)^2$ is much greater than the proton mass squared, $M_p ^2$. In what follows, the effects of non-zero $M_p$ are neglected.
We define light-like components of a four-vector $v^{\mu}$ as $v^{\pm} = v^0 \pm v^z$, where the $Z$-axis is given by the beam direction in the proton-proton c.m.s. From now on we shall use the light-cone coordinates for four-vectors, $v=(v^+,v^-;\vec{v}_T)$, and a perpendicular part $v_{\perp}$ of four-vector $v$ is defined as $v_{\perp} = (0,0;\vec{v}_T)$. Thus, in the light-cone coordinates one has $P_1 = (\sqrt{\sp},0;\vec{0})$, $P_2 = (0,\sqrt{\sp};\vec{0})$, and in what follows the Sudakov decomposition of four-vectors is employed, $v = \alpha_v P_1 + \beta_v P_2 + v_\perp$. In particular, for the DY intermediate electroweak boson one has $q = \alpha_q P_1 + \beta_q P_2 + q_{\perp}$, with $q_{\perp} = (0,0;\vec{q}_T)$, where $\alpha_q$ is equal to the boson Feynman $x_F$, and $\beta_q$ is determined from the mass constraint, $\beta_q = M_T ^2 / x_F \sp$, and the boson transverse mass squared $M_T ^2 = M^2 + q_T ^2$. The boson rapidity in the laboratory frame is $Y = \log{(x_F \sqrt{\sp} / M_T)}$.

In this paper the intermediate boson polarization vectors are defined in the Collins--Soper frame \cite{Collins_Soper_frame} through normalized four-vectors $X^{\mu}$, $Y^{\mu}$ and $Z^{\mu}$ given in the following way (note that differently to the rest of the paper, in the formulas below the $\pm$ labels do not denote the light-cone components of the four vectors):
\be
X^{\mu} = -{M \over \sp q_T M_T} \left( a_{(+)} \tilde P^{\mu} _{(+)} + a_{(-)} \tilde P^{\mu} _{(-)} \right), \qquad 
Z^{\mu} =  {1\over \sp M_T}\left( a_{(-)} \tilde P^{\mu} _{(+)} + a_{(+)} \tilde P^{\mu} _{(-)} \right),
\ee
where $\tilde P^{\mu} _{(\pm)} = P^{\mu} _{(\pm)} - (q\cdot P_{(\pm)} / M^2) q^{\mu}$, $P^{\mu} _{(\pm)} = P^{\mu} _1 \pm P^{\mu} _2$, $a_{(\pm)} = \pm q\cdot P_{(\pm)}$, and $Y^{\mu}$ is defined by orthogonality and normalization conditions: $q\cdot Y = X\cdot Y = Y\cdot Z = 0$, $Y^2 = -1$.
Hence the Collins-Soper polarization four-vectors are:
\be
\epsilon_{(0)} ^{\mu} = Z^{\mu}, \qquad 
\epsilon^{\mu} _{(\pm)} = \mp {1\over \sqrt{2}} (X^{\mu} \pm iY^{\mu}).
\ee

\section{Drell--Yan cross-sections}
\label{Sec:3}

\subsection{Generalities}
\label{Sec:3a}


The Drell--Yan process of lepton--antilepton pair production at the LHC receives contributions from $\gamma^* $ and $Z^0$ or ${Z^0} ^*$ exchange amplitudes. The contributions from   $\gamma^* $ are parity conserving, and from  $Z^0$ include both parity conserving and parity violating terms. In the present paper we do not consider the odd-parity effects coming from the $Z^0$ exchange as they do not affect the Lam--Tung relation breaking and the other observables studied in this paper.

The differential Drell--Yan cross-section in the parity conserving sector may be decomposed into independent angular components (see e.g.\ \cite{ATLASZ0,DYNNLO,Faccioli:2011pn}),
\be
{d\sigma \over dY\, dM^2\, d^2 q_T\, d\Omega_l}  = \sum_{\tau} 
{d\sigma_{\tau} \over dY\, dM^2 \, d^2 q_T}\, g_{\tau}(\Omega_l), \quad
\Omega_l = (\theta ,\phi), \quad \tau \in \{L,T,TT,LT\},
\ee
where
\begin{eqnarray}
d\sigma_L  =  {\cal C}_L \;    d\sigma^H _{00}, \nonumber  & \hspace{5ex} & 
d\sigma_T  =  {\cal C}_L \;    {d\sigma^H _{++} + d\sigma^H _{--} \over 2},  \nonumber \\
d\sigma_{TT} =  {\cal C}_L \; {d\sigma^H _{+-} + d\sigma^H _{-+} \over 2}, & &
d\sigma_{LT} = {\cal C}_L \; {d\sigma^H _{+0} + d\sigma^H _{0+} - d\sigma^H _{-0} - d\sigma^H _{0-} \over 2\sqrt{2}},  
\end{eqnarray}
where  $d\sigma^H _{\sigma\sigma'}$ are the hadronic cross-sections\footnote{More adequately, $d\sigma^H _{\sigma\sigma'}$ are the cross-section only for $\sigma=\sigma'$, and are proportional to density matrix elements for the boson production in the boson helicity basis when the initial and final boson helicities, $\sigma$ and $\sigma'$ are different.} for the electroweak boson production in the helicity basis, and the proportionality constant ${\cal C}_L = {\alphaem M^2 \over 8\pi^2} |D_V(M^2)|^2$ accounts for the boson exchange amplitude and the normalization of the leptonic part of the amplitude. The scalar part of the boson propagator with $q^2 = M^2$ reads $D_V(M^2) = 1/(M^2 - M_V^2 + i\Gamma_V M_V)$, where $M_V$ and $\Gamma_V$ are the boson~$V$ mass and decay width respectively. The functions of the lepton angles $(\theta,\phi)$ in the lepton pair c.m.s.\ read 
\begin{eqnarray}
g_L(\Omega_l) = 1 - \cos^2 \theta ,  &\hspace{5ex} & 
g_T(\Omega_l)  =  1 + \cos^2 \theta ,  \nonumber\\
g_{TT}(\Omega_l)  =  \sin ^2 \theta \, \cos 2\phi, & &
g_{LT}(\Omega_l)  =  \sin 2\theta \, \cos \phi .
\end{eqnarray}
It is customary to parameterize the Drell--Yan lepton angular distributions exchange in terms of harmonic functions with $A_i$ coefficients in the following way: 
\begin{eqnarray}
\left[ {d\sigma \over dY\, dM^2\, d^2 q_T} \right]^{-1}\,  
\frac{d\sigma}{ dY\, dM^2\, d^2 q_T\, d\Omega_l}  & = & 
{3\over 16\pi} \left[ (1 + \cos^2 \theta) + {1\over 2} A_0 \,(1 - 3\cos^2 \theta) \right. \nonumber \\
& + &  \left. A_1\, \sin 2\theta \cos \phi + {1\over 2} A_2 \, \sin^2 \cos 2 \phi \right],
\end{eqnarray}
where the odd-parity effects coming from the $Z^0$ exchange are neglected. It is straightforward to express the coefficients $A_i$ through the boson hadroproduction cross-sections $d\sigma_{\tau}$. The results read:
\be
A_0 = {\sigma_L \over \sigma_T + \sigma_L / 2}, \quad A_1 = {\sigma_{LT} \over \sigma_T + \sigma_L / 2}, \quad A_2 = {2\sigma_{TT}\over \sigma_T + \sigma_L / 2},
\ee
where $\sigma_{\tau}$ stands for suitably integrated $d\sigma_{\tau}$.

Note that in the parity conserving part of the Drell--Yan cross-section the $Z^0$~boson hadroproduction cross-sections are proportional to the $\gamma^*$~hadroproduction cross-sections.
At the amplitude level, the photon coupling $ee_f \gamma_{\mu}$ to a quark of flavor $f$ is proportional to the quark charge $e_f$, while the quark coupling to the $Z^0$ boson is $g_V ^f \gamma_{\mu} - g_A ^f \gamma_{\mu}\gamma_5$ where $g_V ^f$ and $g_A ^f$ are the vector and axial-vector quark couplings to the $Z^0$ boson. In the parity even cross-sections the $\gamma_5$ may be anticommuted and eliminated from the amplitudes squared and the resulting expressions are proportional to the virtual photon production cross-sections, with a substitution $(e e_f)^2 \to (g^ f _A) ^2 + (g_V ^f)^2$.

\subsection{Approximations applied}
\label{Sec:3b}

\paragraph{The high energy limit and the $k_T$-factorization approach.} 
In this paper we describe the Drell--Yan amplitudes within the $k_T$-factorization (or high energy factorization) framework, in which the high-energy limit is employed. This limit combined with the (fixed order) perturbative expansion of scattering amplitudes allows for a systematic selection of leading diagrams that contribute to the scattering amplitudes. The key feature of this approach is explicit inclusion of the initial parton transverse momenta and virtualities.  

The typical value of quark--parton~$x_q$ in Drell--Yan processes at the LHC at $\sqrt{\sp} = 8$~TeV is $x_q \sim M_T / \sqrt{\sp} < 0.02$  for masses up to the $Z^0$ peak and the boson transverse momentum $q_T < 100$~GeV. In this range of $x_q$ the quark distributions are dominated by the sea quarks, and for $pp$ collisions antiquarks come entirely from the sea. Due to the spin of $1/2$ the exchange of quarks and antiquarks in the $t$-channel decreases exponentially with the exchange rapidity span, while the spin~1 gluon exchange amplitude grows with the rapidity span.  This feature combined motivates the standard approximation in which the sea parton evolution is driven by the evolution of the exchanged gluon density, and the sea quarks that enter the hard matrix element are generated in the matrix element, or in the last step of parton evolution as a result of the gluon splitting. In our analysis we adopt this approximation. 

\begin{figure}
\centering
\begin{tabular}{lll}
\parbox{0.35\columnwidth}{\includegraphics[height=.4\textwidth]{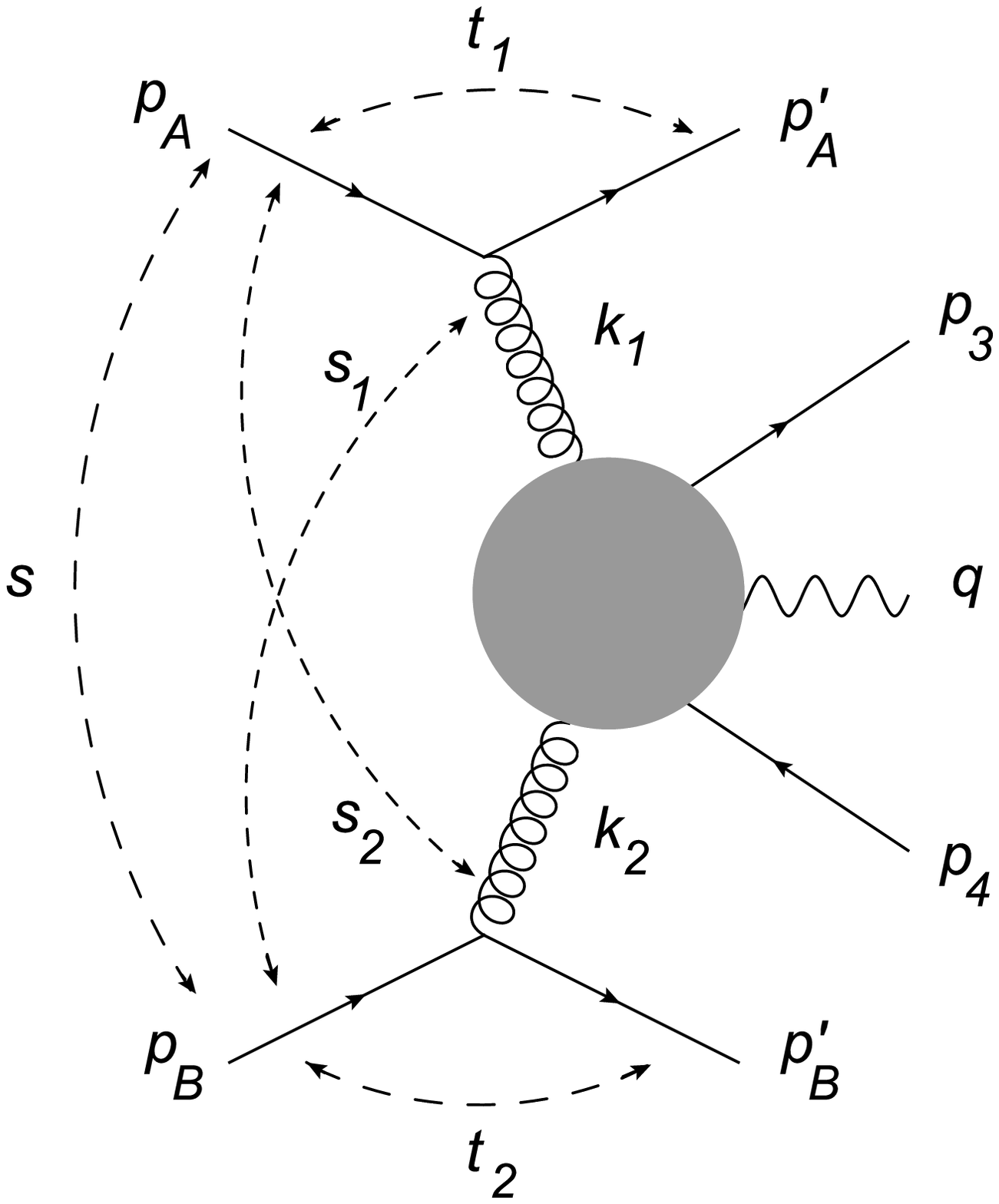}} \hspace{1em} &
\parbox{0.24\columnwidth}{\includegraphics[height=.28\textwidth]{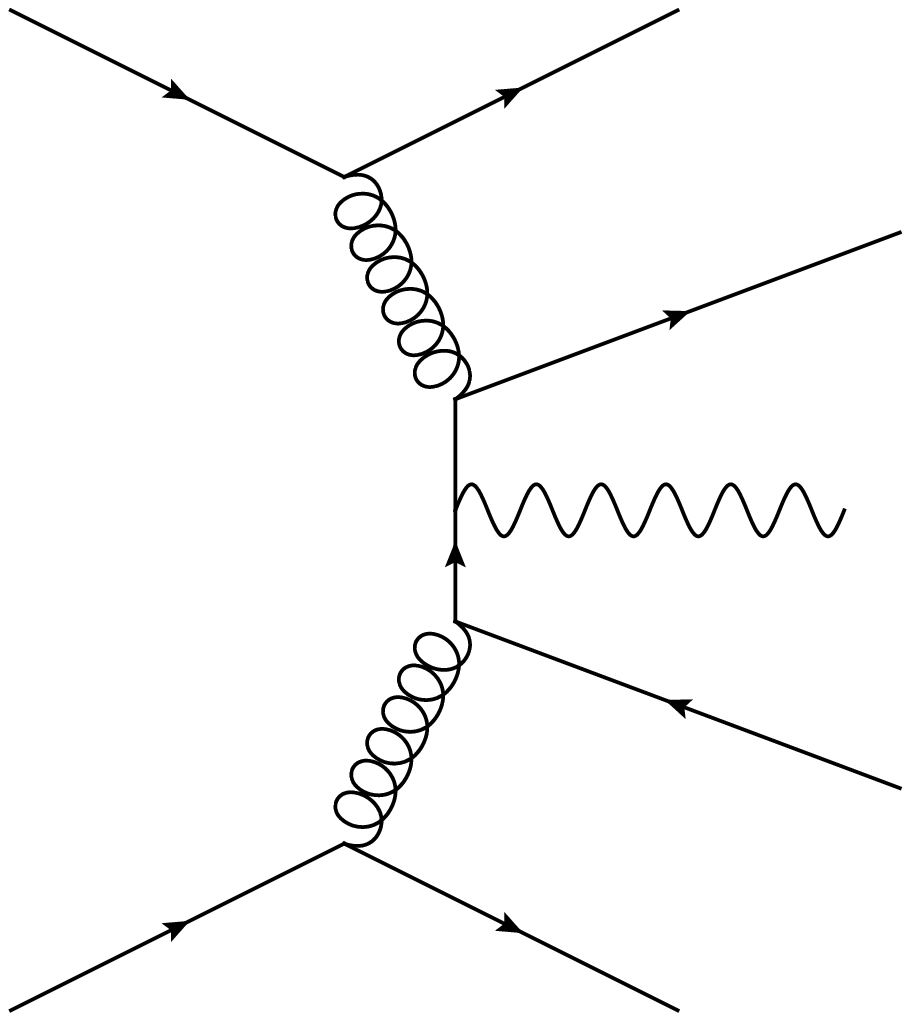}} \hspace{1em} &
\parbox{0.24\columnwidth}{\includegraphics[height=.28\textwidth]{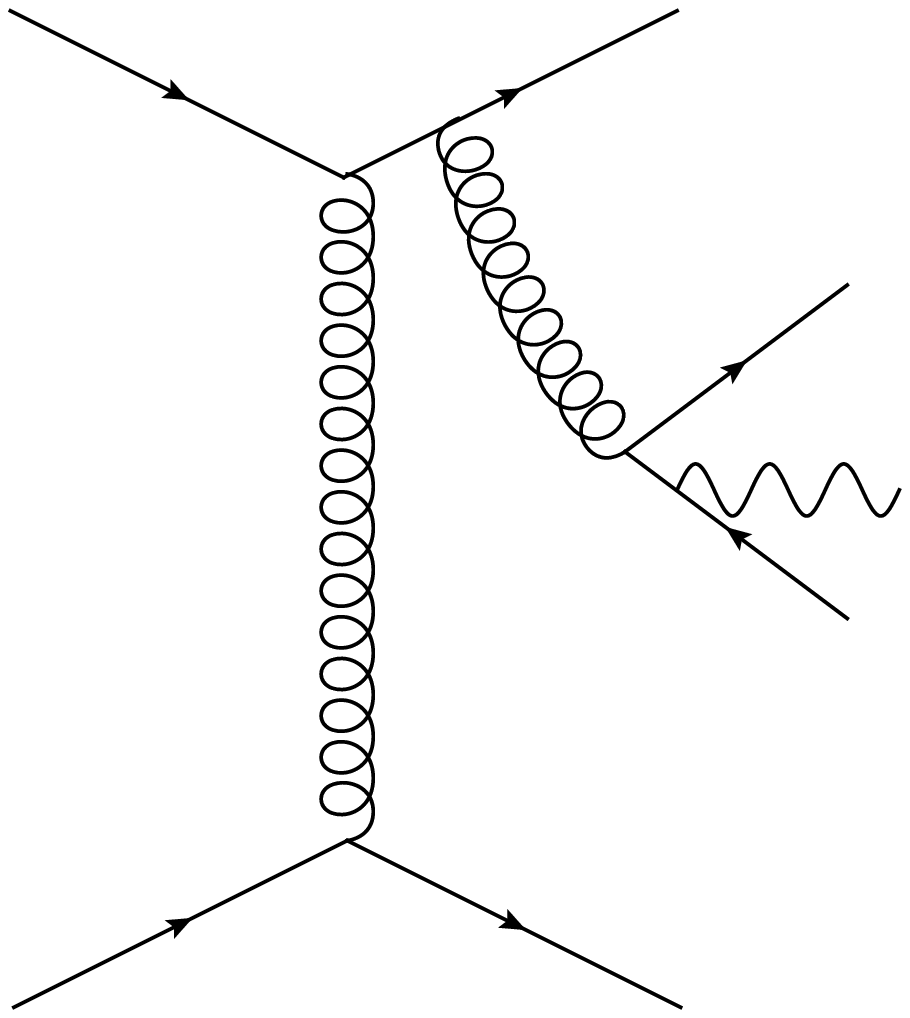}} \\
{\Large a)}& {\Large b)} & {\Large c)}
\end{tabular}
\caption{Diagrams illustrating discussion of the $g^*g^*$ channel in the Drell--Yan process:
a) embedding of the $g^*g^*$ channel into scattering of two fast quarks, four-momenta of particles and selected Mandelstam invariants are indicated; b) an example of a diagram with explicitly factorizing topology;  c) an example of diagram that apparently breaks factorization (see the text for explanations).  
\label{Fig:ofshellampl}}
\end{figure}


Below we discuss in some detail approximations made in the $k_T$-factorization framework and consistency of the obtained formulae for the contribution of two gluons with non-zero $k_T$ (i.e.\ the $g^*g^*$ contribution) to the Drell-Yan scattering. A thorough analysis of this channel within the $k_T$-factorization framework was performed in Refs.\ \cite{DeakSchwen,BLZdy}. For completeness of the presentation we shortly quote below the main steps and results obtained in these papers, and discuss the necessary kinematic conditions for the approximation to hold. The framework and the approximations applied are based on the classical papers \cite{CaCiaHaut}.

A particularly convenient starting point for such an analysis was proposed in Ref.\ \cite{DeakSchwen}, where the electroweak boson $V$ production in association with heavy $b\bar b$ quark pair in the $g^* g^*$ channel is analyzed. In this reference a tree level topology for scattering of two light quarks, $q_A$ and $q_B$, with four-momenta $p_A$ and $p_B$ correspondingly, into $q_A(p_A ') q_B (p_B ') b(p_3)\bar b(p_4) V(q)$  final state is considered, see Fig.\ \ref{Fig:ofshellampl}a. For the light quarks one assumes the zero mass approximation, $p_A^2 = p_B^2 = {p_A'} ^2 = {p_B'}^2 = 0$. At the lowest order in QCD this process occurs via exchange of two virtual gluons between the scattering light quarks and the produced heavy $b$-quarks. The scattering amplitude is evaluated in the standard way with all Feynman diagrams at ${\cal O}(\alpha_s^2)$ in QCD, hence it is gauge invariant.

Next, one uses the high energy approximation to separate the amplitude of virtual gluon emission from the light quarks from the $g^* g^* \to b\bar b V$ scattering amplitude. This is done in the following way: the virtual gluon momenta, $k_1$ and $k_2$, are written in terms of Sudakov variables, $k_i = \alpha_i p_A + \beta_i p_B + k_{i \perp}$, where $k_{i \perp}$ stands for the gluon $i$ four-momentum component transverse to the plane spanned by $p_A$ and $p_B$.
In what follows, we shall also use the transverse momenta, $\vec{k}_i$ corresponding to $k_{i \perp}$.

It is convenient to introduce the Mandelstam variables: 
\be
s = (p_A + p_B)^2, \qquad s_1 = (p_B + k_1)^2, \qquad s_2 = (p_A + k_2)^2, 
\ee
and
\be
t_1 = (p' _A - p_A)^2 = k_1 ^2, \qquad t_2 = (p' _B - p_B)^2 = k_2^2.
\ee
Clearly, the bulk of the cross section comes from the region of strong hierarchy of the invariants $s_i \gg |t_j|$, $i,j=1,2$. This hierarchy follows from the fact that transverse momenta of the scattered light quarks are much smaller than the c.m.s.\ energy of the process and the invariant masses, $\sqrt{s_i}$, of the gluon-light quark pairs. We do not assume a strong hierarchy of $s_i$ and $s$.

The straightforward kinematic analysis of the scattering process gives the following results for the 
Sudakov coefficients $\alpha_i$ and $\beta_i$: 
\be
\alpha_1 = {s_1 + \vec{k}_1^2 \over (1+ \beta_1)s}, \qquad \beta_1 = -{\vec{k}_1 ^2 \over s(1-\alpha_1)},  
\qquad
\alpha_2 = -{\vec{k}_2^2 \over (1- \beta_2)s}, \qquad \beta_2 = {s_2 + \vec{k}_2 ^2 \over s(1+\alpha_2)}. 
\ee
Taking into account that $t_i \ll s_j$, we obtain the following approximate expressions:
\be 
\alpha_1 \simeq {s_1 \over s}, \qquad \beta_1 \simeq {t_1 \over s (1-\alpha_1)}, \qquad
\alpha_2 \simeq {t_2 \over s (1-\beta_2)}, \qquad \beta_2 \simeq {s_2 \over s},
\ee
that hold up to ${\cal O}(t_i / s_j)$ accuracy. Although $\alpha_1$ and $\beta_2$ are not necessarily small, the regions of $\alpha_1, \beta_2 \to 1$ where two of the denominators are close to zero are strongly suppressed in the cross sections due to strong suppression of parton distributions for $\alpha_1, \beta_2 \to 1$. It follows that $\alpha_1 \sim \beta_2 \gg \alpha_2 \sim \beta_1$ and $t_1 = k_1 ^2 \simeq -\vec{k}_1 ^2$,  $t_2 = k_2 ^2 \simeq -\vec{k}_2 ^2$. Hence in the high energy approximation one neglects $\beta_1$ and $\alpha_2$. In the explicit numerical analysis performed in the next sections we checked that in the kinematic conditions of the $Z^0$ production at the LHC one probes typically $s_i \sim 1$~TeV$^2$ and $|\vec k_i| \sim 0.1$~TeV, and the approximation parameter is small, $|t_i/s_j| \sim 0.01$, thus this approximation is justified.
Note also, that in the collinear approximation for the gluons one sets $|\vec k_i| = 0$, then $\beta_1 = \alpha_2 = 0$ and $\alpha_1$ and $\beta_2$ are identified with the gluon--parton $x$ variables, $x_1 = \alpha_1$, $x_2 = \beta_2$. 

After the kinematic hierarchy and approximations are established, one turns to the coupling of gluons to the scattering light quarks. The amplitude of the gluon coupling to quark $q_A$ is proportional to 
$ -i \bar u_{\lambda _A '}(p_A ') \gamma^{\mu} u_{\lambda_A}(p_A)$, 
Where $u_{\lambda _A}(p_A)$ and $u_{\lambda _A '}(p_A ')$ denote the spinors of the incoming and outgoing quark with helicity $\lambda_A$ and $\lambda_A '$ correspondingly. 
In the high energy approximation described above, one gets
\be 
 -i \bar u_{\lambda _A '}(p_A ') \gamma^{\mu} u_{\lambda_A}(p_A) = -2i p_A ^{\mu} \, \delta_{\lambda _A ', \lambda_A} + \ldots
\ee
where the neglected terms denoted by `$\ldots$' are power supressed in the kinematic small $\beta_1$ expansion.
This motivates the eikonal approximation for the virtual gluon polarization assumed in the high energy limit, see Sec.\ \ref{Sec:3c} for more details and the explicit fomulae. Note, that for this approximation to be valid it is not necessary to assume that $x_1 = \alpha_1$ of the gluon is small.

The above analysis is carried out for the process initiated by the light quark scattering, and the discussed amplitude is gauge invariant by construction. In order to complete the description of the process in the $k_T$-factorization framework it is necessary to factorize the gluon coupling to incoming quarks / hadrons from the off-shell hard matrix element describing the $g^* g^* \to b \bar b V$ process. In an arbitrary gauge the factorization may be not obvious, as there are contributions to the process that correspond to diagram topologies of a direct gluon exchange between the scattering hadrons accompanied by a virtual gluon emission followed by a splitting $g^* \to b\bar b V$ as illustrated in Fig.\ \ref{Fig:ofshellampl}c, see e.g.\ Ref.\ \cite{DeakSchwen} for more details. Such contributions are, however, not independent of the remaining diagrams corresponding to the standard $g^* g^*$ scattering (as illustrated in Fig.\ \ref{Fig:ofshellampl}b) --- the amplitudes of the different topologies are related by the gauge invariance constraint. The emerging factorization can be seen in two ways: (i) one may work in an axial (hence physical) gauge, in which the apparent non-factorizing contributions explicitly vanish, or (ii) combine all the diagrams in an arbitrary gauge and show that the apparently non-factorizing diagrams may be absorbed into an universal, gauge invariant effective triple virtual gluon vertex, that explicitly obeys Ward identities. Both the approaches are described extensively in the literature, see e.g.\  \cite{CaCiaHaut,DeakSchwen,BLZdy}. It is important to stress that, again, in this procedure one adopts the high energy approximation relying on the condition of small $\beta_1$ and $\alpha_2$, and it is not necessary to assume that gluon $x$-es are small. In the present analysis we use the latter approach, incorporating the gauge invariant effective triple gluon vertex $V_{\mathrm{eff}}$ (see Appendix \ref{App:2} for the explicit form), and the gluon propagators are taken in the Feynman gauge. 

So far we followed the process selection of Ref.\ \cite{DeakSchwen} --- the electroweak boson production with association with the heavy $b\bar b$ quark pair. This is particularly convenient for studies of the $g^*g^*$ partonic channel, as the $b$ quark partons in the proton may be safely assumed to come only from gluon splittings, and the $g^* g^*$ channel contributions exhaust the cross-section. The logic applied here closely follows the classical approach for heavy quark production in the high energy approximation \cite{CaCiaHaut}. 
If, instead of the heavy quarks $b\bar b$ the light quarks are produced in association with the electroweak boson, the evaluation of the $g^*g^*$ channel contribution is exactly the same (see e.g.\ Refs.\ \cite{BLZprompt,BLZdy}), however for the light quarks in the final state, additional initial $q\bar q$, $g q$ and $g \bar q$  parton channel contributions are included.


\paragraph{Relation of the chosen scheme to collinear factorization approach.}
It may be useful for more clarity to discuss the connection between the high energy limit approach described above and the collinear factorization framework. In the collinear approximation the following channels contribute to the neutral electroweak (real or virtual) boson $V$ production in $pp$ collisions:
\begin{itemize}
\item from the LO, ${\cal O}(\alphaem)$: $q\bar q \to V$;
\item from  the NLO, ${\cal O}(\alphaem\alpha_s)$: $qg \to Vq$, $\bar q g \to V \bar q$, $q\bar q \to V g$;   
\item from the NNLO, ${\cal O}(\alphaem\alpha_s^2)$: $q\bar q \to V g g$, $q \bar q \to  q \bar q V$, $q q \to q q V$, $\bar q \bar q \to \bar q \bar q V$, $q g \to q g V$, $\bar q g \to \bar q g V$ and $gg \to q\bar q V$. 
\end{itemize}

In the above partonic channels one may treat separately contributions of the valence quarks, $\qval$ and of the
sea quarks, $\qsea$, $\bqsea$. In the high energy approximation described above the sea (anti-)quarks enter only in the hard matrix element, so from the partonic channels up to the NNLO, one is left with contributions of $\qval g \to q V$, $\qval  g \to q g V$, $\qval  \qval  \to q q V$, and $gg \to q\bar q V$.
In the high energy limit the channel $\qval  \qval  \to q q V$ is driven by a gluon exchange between the valence quarks and is absorbed  into the $\qval g \to q V$ channel, as the gluon distribution contains gluon emissions from valence quarks, see the discussion above. The contribution of $\qval g \to q V$ to the boson production cross section carries the first power of $\alphas $, and both $\qval  g \to q g V$ and $gg \to q\bar q V$ enter as ${\cal O}(\alphas ^2)$ contributions in QCD. 

In the relevant kinematic regime the quark--parton $x_q \sim 0.01$ in the electroweak boson emission vertex, however, due to the gluon and sea quark distributions being much greater than the distribution of the valence quarks, the contribution of $gg \to q\bar q V$ is expected to be significantly larger than $\qval  g \to q g V$ so we neglect the latter. Thus, the two dominant channels in the high energy limit are: $\qval  g \to q V$ (Fig.\ \ref{Fig:channels}b)), and $gg \to q\bar q V$ (Fig.\ \ref{Fig:channels}c)). In our approach, which includes the treatment of non-zero gluon transverse momentum $k_T$, the incoming gluons, $g^*$, are virtual, and the $g^*$ density in the proton will be parameterized in terms of the gluon Transverse Momentum Distribution ${\cal F}(x,k_T ^2,\mu_F)$. The valence quark distributions are expected to have much narrower distribution in the transverse momentum, so for the valence quarks we keep the collinear approximation, which is viewed in our approach as the small $k_T$-width limit of the corresponding quark TMD.


The partonic diagram selection based on the gluon exchange dominance in the high-energy limit may be related to the rigorous systematic expansion of the collinear approach. Thus, the $\qval  g^* \to q V$ hard matrix element contains the LO DGLAP $g \to \bar q$ splitting combined with the LO $\qval \bqsea \to V$ matrix element, the contribution of the NLO $\qval g\to qV$ term and contributions beyond the collinear NLO level that emerge because of the exact treatment of parton kinematics. On the other hand, the loop corrections to $q\bar q \to V$ amplitude are neglected, that also contribute to collinear NLO approximation. For the $g^* g^* \to q\bar q V$ channel, the hard matrix element accounts for the collinear LO $\qsea\bqsea \to V$ contribution preceded by $g\to \qsea$ and $g \to \bqsea$ DGLAP splittings, the NLO contributions $\qsea g \to q V$ and $\bqsea g \to \bar q V$, preceded by $g\to \qsea$ or $g \to \bqsea$ DGLAP splittings, the leading contribution to the NNLO collinear matrix element $gg \to q\bar q V$, and some contributions beyond the DGLAP NNLO coming from the exact treatment of parton scattering kinematics. Again, loop corrections are not included in this treatment. Hence our approximation is expected to cover the leading contributions to the collinear partonic channels $\qval \bqsea \to V$,  $\qsea\bqsea \to V$, $\qval  g \to q V$, $\qsea g \to q V$, $\bqsea g \to \bar q V$ and $gg \to q\bar q V$. These channels are dominant ones in the neutral electroweak boson production, and certain loss of completeness w.r.t.\ the existing collinear NNLO calculation is justified by more accurate treatment of parton scattering kinematics in the $k_T$-factorization approach.

\subsection{Drell--Yan cross-sections in the partonic channels}
\label{Sec:3c}

\begin{figure}
\centering
\begin{tabular}{lll}
\includegraphics[height=.22\textwidth]{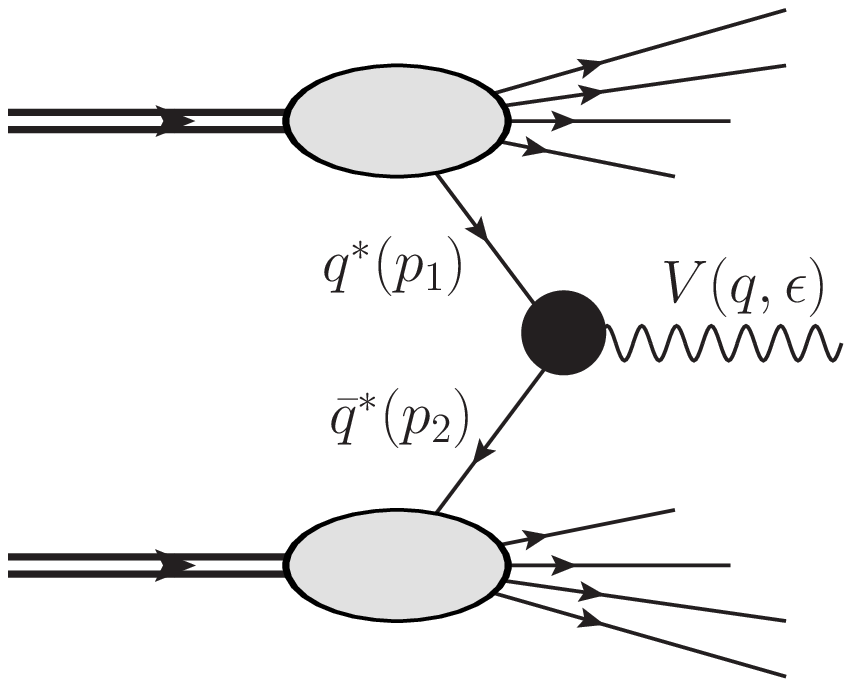} \hspace{1em} &
\includegraphics[height=.22\textwidth]{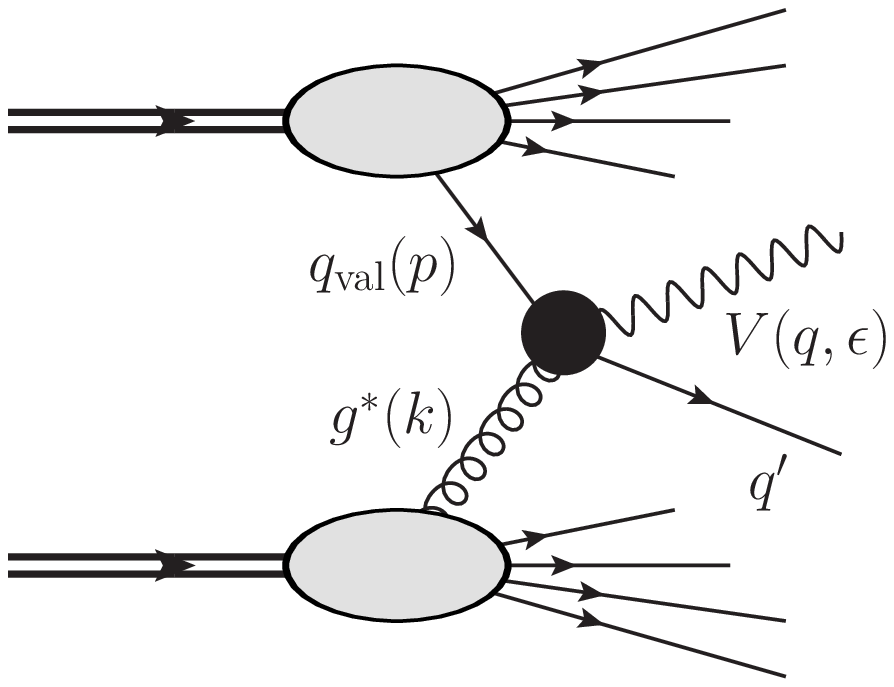} \hspace{1em} &
\includegraphics[height=.22\textwidth]{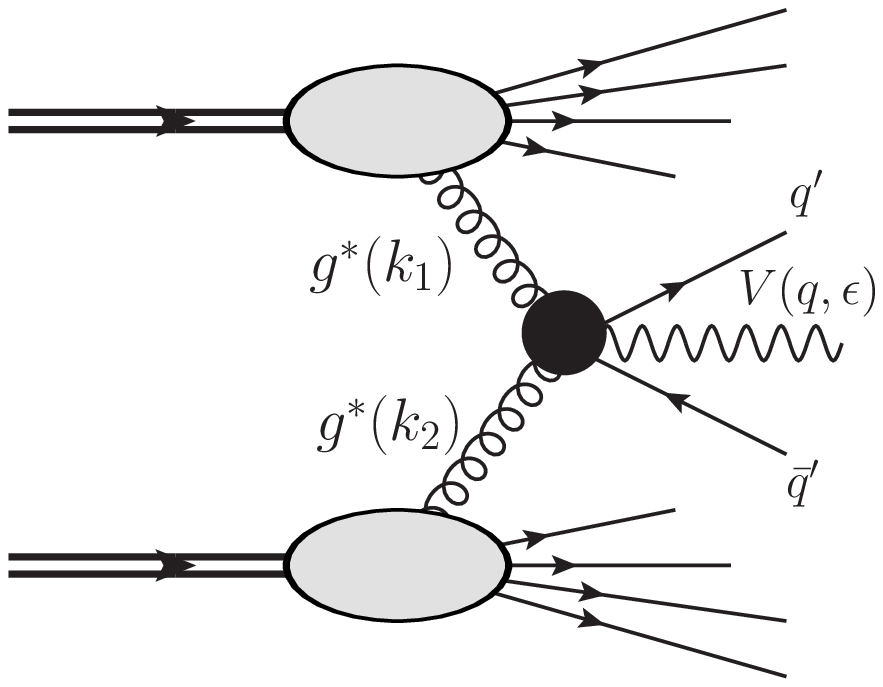} \\
{\Large a)} & {\Large b)} & {\Large c)} 
\end{tabular}
\caption{\it Partonic channels in the electroweak boson ($V=\gamma^*$, $Z^0$ or ${Z^0} ^*$) hadroproduction: a) the $q^* {\bar q}^* \to V$ channel;  b) the $\qval g^* \to qV$ channel and c) the $g^* g^* \to q\bar q V$ channel. The black blobs represent channel dependent effective vertices. 
\label{Fig:channels}}
\end{figure}

With the approximations described above the Drell--Yan cross-sections may be expressed as sums of the
corresponding contributions from the $\qval g^*$ channel and the $g^*g^*$ channel,
\be
 d\sigma^H _{\sigma\sigma'} = d\sigma^{(qg^*)} _{\sigma\sigma'} +  d\sigma^{(g^*g^*)} _{\sigma\sigma'}.
\ee
see Fig.\ \ref{Fig:channels}b, c.
The quark--gluon channel contribution in the $k_T$-factorization approach was derived in Ref.\ \cite{Brodsky,Kopeliovich,GJ} (see also Refs.\cite{MSS,SchSz}). The original calculation was performed in the 
 Gottfried--Jackson frame \cite{Gottfried:1964nx} (alternatively called the $u$-channel helicity frame \cite{Brodsky}). The explicit expression reads, 
%
%
\begin{eqnarray}
\frac{\tHi _{\sigma\sigma'}}{d Y dM^2 d^2 q_T}&=& 
\sum_{f}
\int_{x_F}^1 dx_q \ \wp_{f,\mathrm{val}}(x_q,\mu_F) 
\int {d^2 \vec{k}_T \over \pi \vec{k}_T ^2} \nonumber \\
& & \times \; \frac{4\pi \alphas (\mu_F)}{3} \; 
{\cal F}({x}_g,\vec{k}_T^2,\mu_F) \;
\tilde{\Phi}^{(f)} _{\sigma \sigma'} (\vec{q}_T,\vec{k}_T,z=x_F/x_q),
\label{dsigma2b}
\end{eqnarray}
where $x_g = (  (1-z)M^2 + q_T^2 )/ ( \sp x_F (1-z) )$, $\lambda_1$ and $\lambda_2$ are helicities of the incoming and outgoing quark, the index $f$ runs over quark flavors, $f \in \{d,u,s,c,b\}$, $\wp_{f,\mathrm{val}}$ is the collinear valence quark~$f$ distribution function, and all the quarks are assumed to be massless when compared to the DY pair mass $M$. The helicity dependent $\gamma^*$ impact factors are
\begin{eqnarray}
\tilde{\Phi}^{(f)} _{\sigma \sigma'} (\vec{q}_T,\vec{k}_T,z)= e_f^2 \sum_{\lambda_1,\lambda_2=+,-} \ {A_{\lambda_1,\lambda_2}^{(\sigma)}}^\dagger A_{\lambda_1,\lambda_2}^{(\sigma')},
\label{formfactor}
\end{eqnarray}
where with the chosen set of $\gamma^*$ polarization vectors, 
\begin{eqnarray}
A_{\lambda_1,\lambda_2}^{(0)} &=& \frac{e}{4\pi} \delta_{\lambda_1,\lambda_2} 
\left[ \frac{M(1-z)}{M^2(1-z)+ \vec{q}_T^{\ 2}} - \frac{M(1-z)}{M^2(1-z)+ (\vec{q}_T-z\vec{k}_T)^2 }  \right]
\label{amplitudesMom0} ,
\end{eqnarray}
\begin{eqnarray}
A_{\lambda_1,\lambda_2}^{(\pm)} &=& \frac{e}{8\pi} 
\delta_{\lambda_1,\lambda_2}(2-z \mp \lambda_1 z) \nonumber \\
& & \times \left[ \frac{- \vec{q}_T }{M^2(1-z)+\vec{q}_T^{\ 2}} - \frac{-(\vec{q}_T-z\vec{k}_T) }{M^2(1-z)+(\vec{q}_T-z\vec{k}_T)^2 }  \right] \cdot \vec{\epsilon}_\bot^{\ (\pm)} ,
\label{amplitudesMom+}
\end{eqnarray}
are proportional to the DY $\gamma^*$ emission amplitudes. Note that we suppressed the arguments  $\vec{q}_T$, $z$ and $\vec{k}_T$ of $A_{\lambda_1,\lambda_2}^{(\sigma)}$. 

Next, the polarization vectors are transformed from the $u$-channel helicity frame to the Collins-Soper frame \cite{Collins_Soper_frame}. 
This results with linear transformation of the cross-sections in the helicity basis that may be written as,
\be
\Hi_{\tau} = \sum_{\tau'}\, {\cal R}_{\tau\tau'} \, \tHi_{\tau'}, \qquad \tau,\tau' \in {L,T,TT,LT}.
\ee
The explicit form of the transformation represented by ${\cal R}_{\tau\tau'}$ is given in Appendix \ref{App:1}.


\begin{figure}
\centering
\includegraphics[width=.18\textwidth]{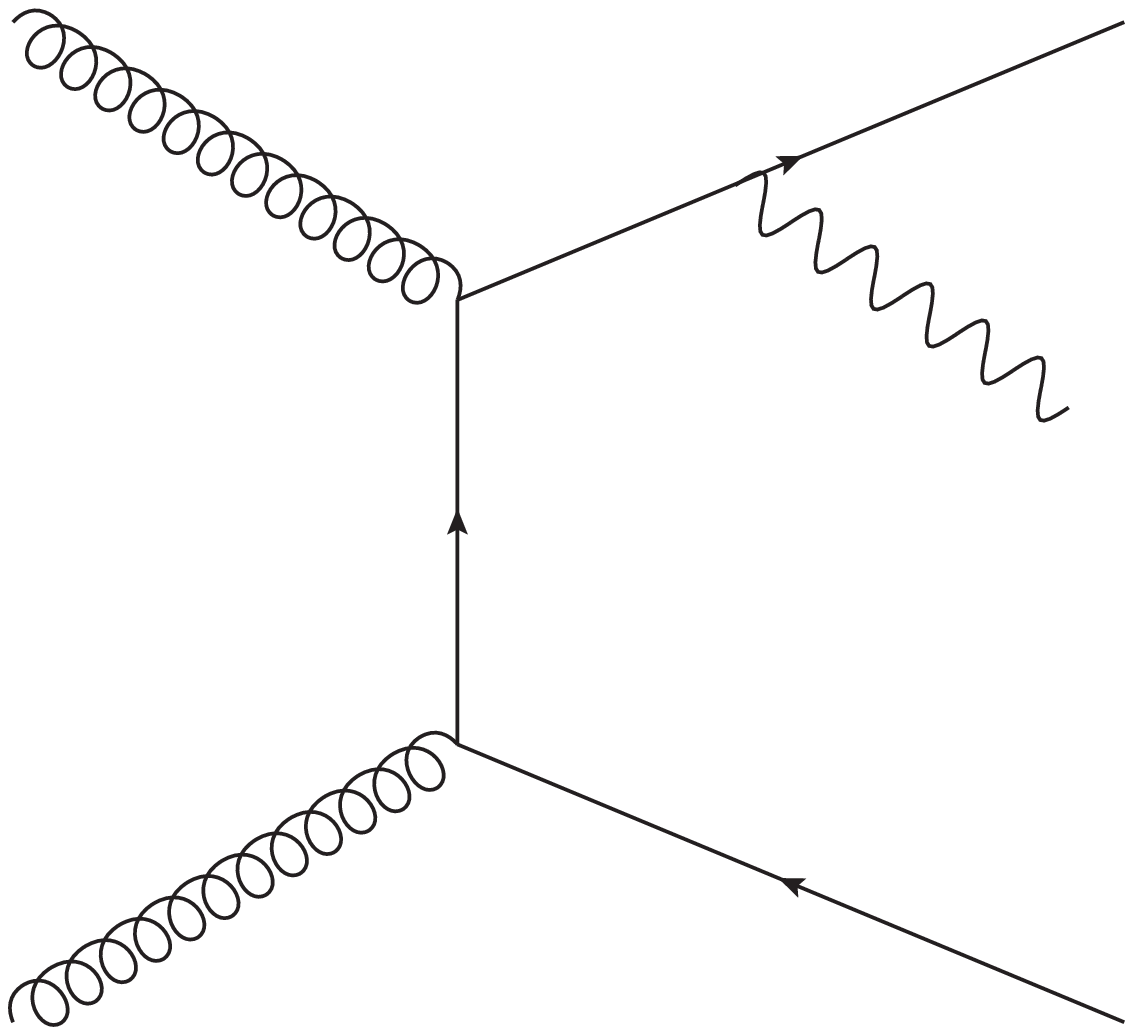} \hspace{5em}
\includegraphics[width=.18\textwidth]{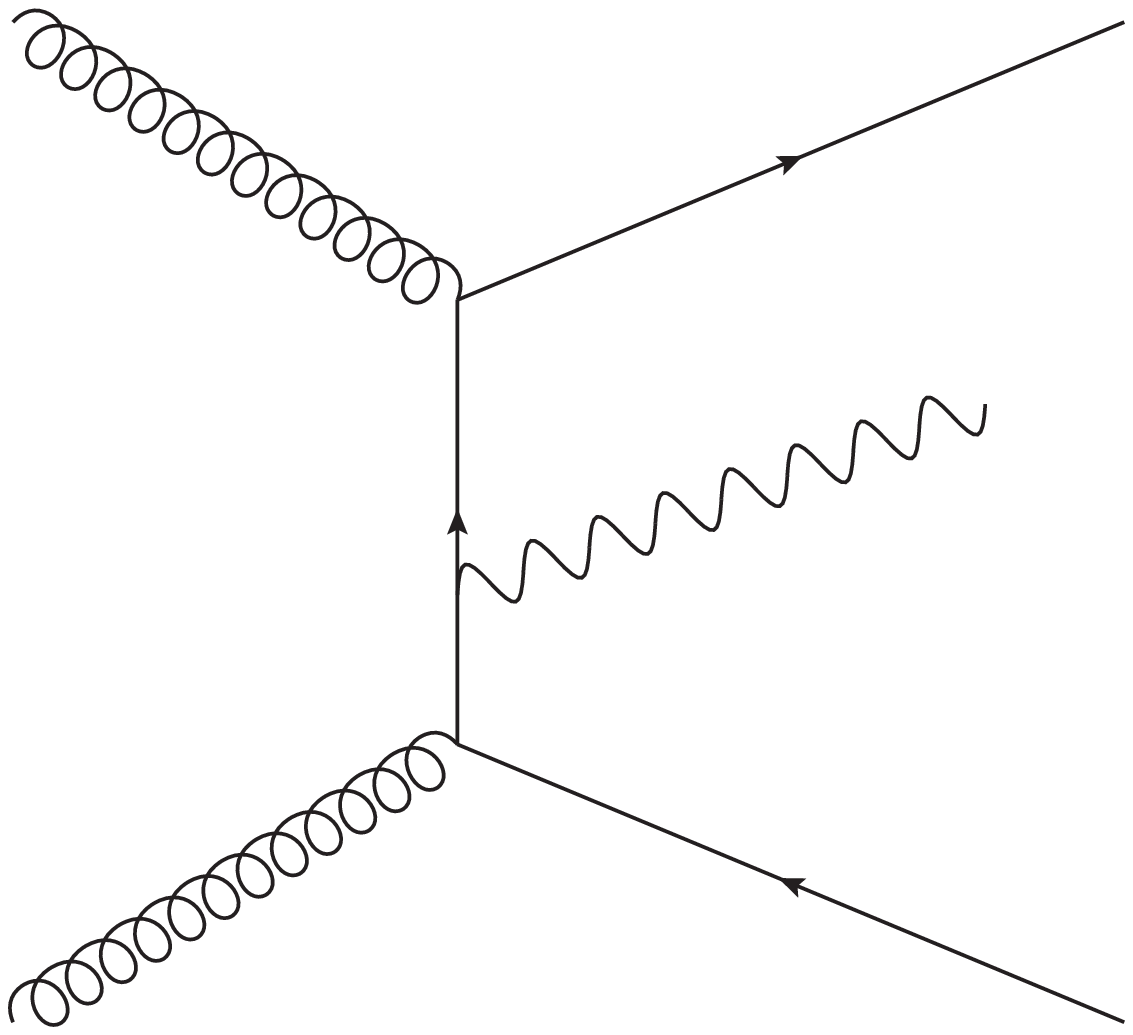} \hspace{5em}
\includegraphics[width=.18\textwidth]{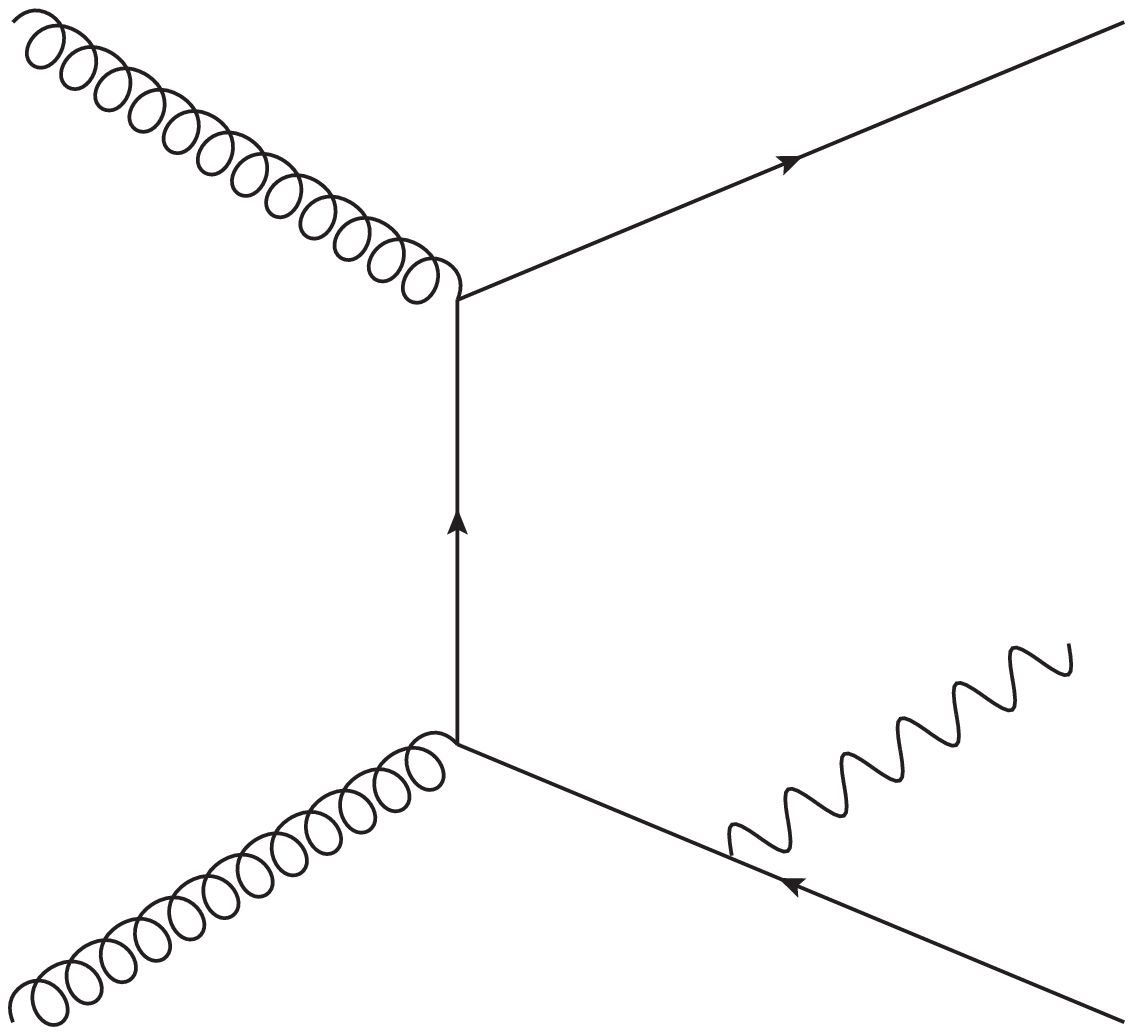} \\[1em] 
\includegraphics[width=.18\textwidth]{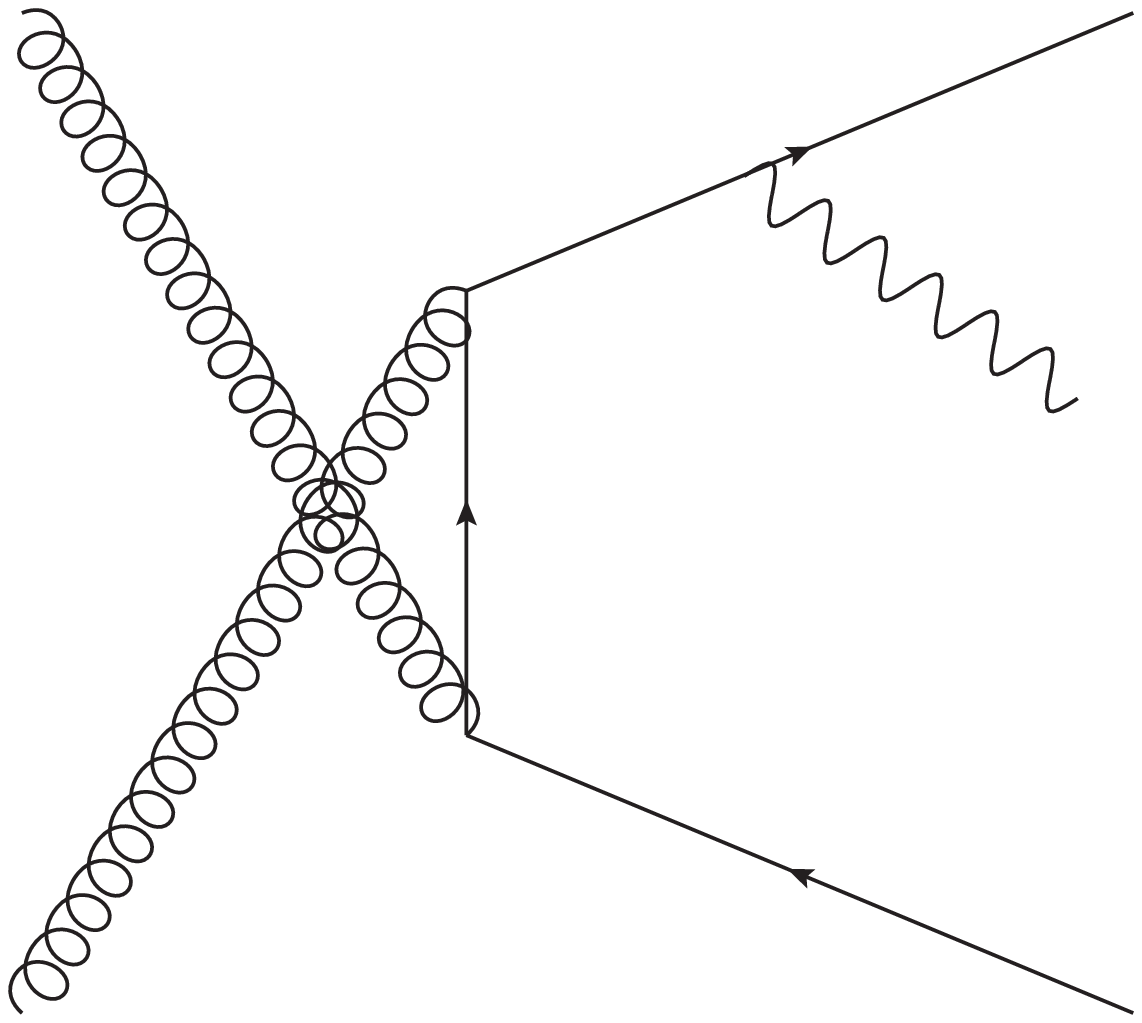} \hspace{5em}
\includegraphics[width=.18\textwidth]{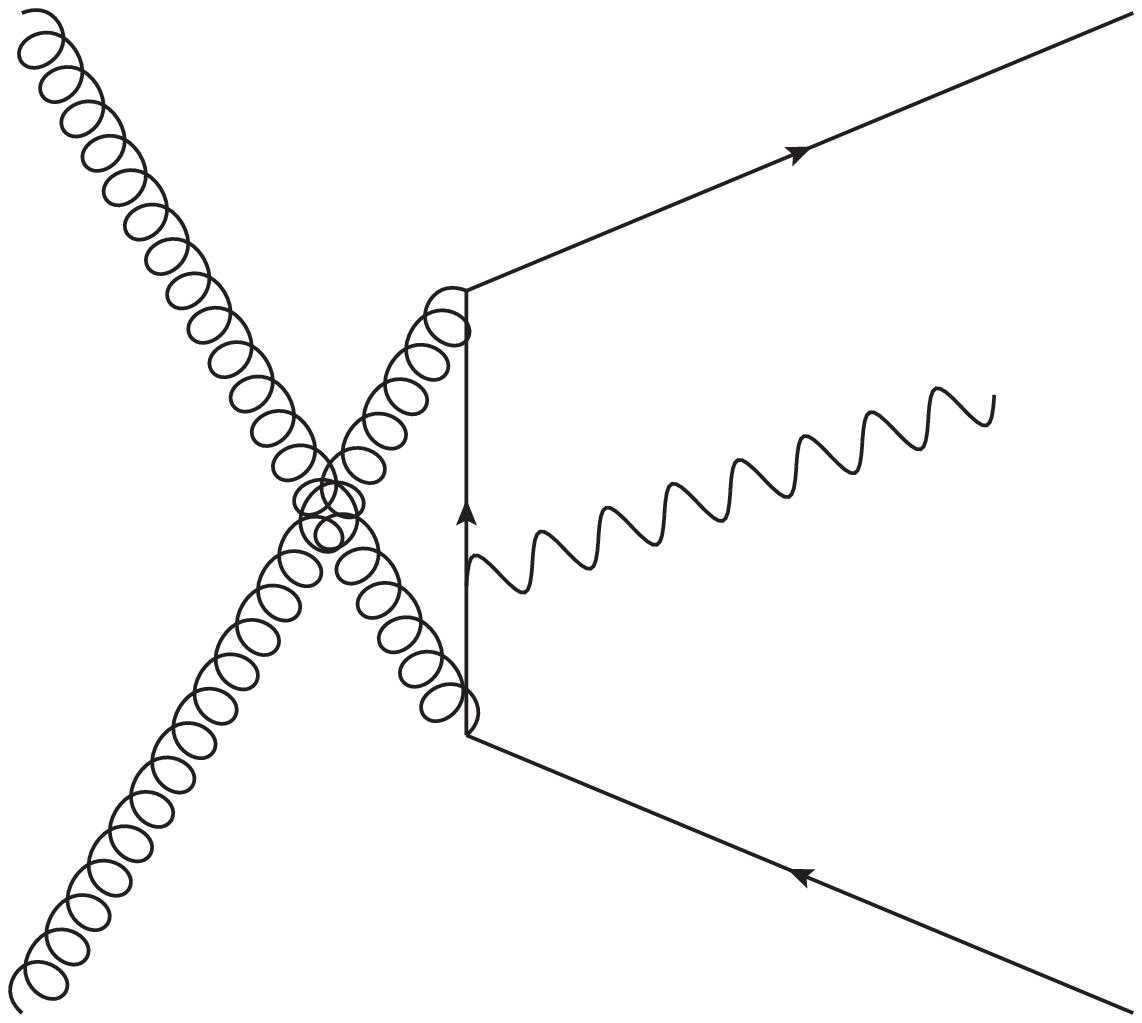} \hspace{5em} 
\includegraphics[width=.18\textwidth]{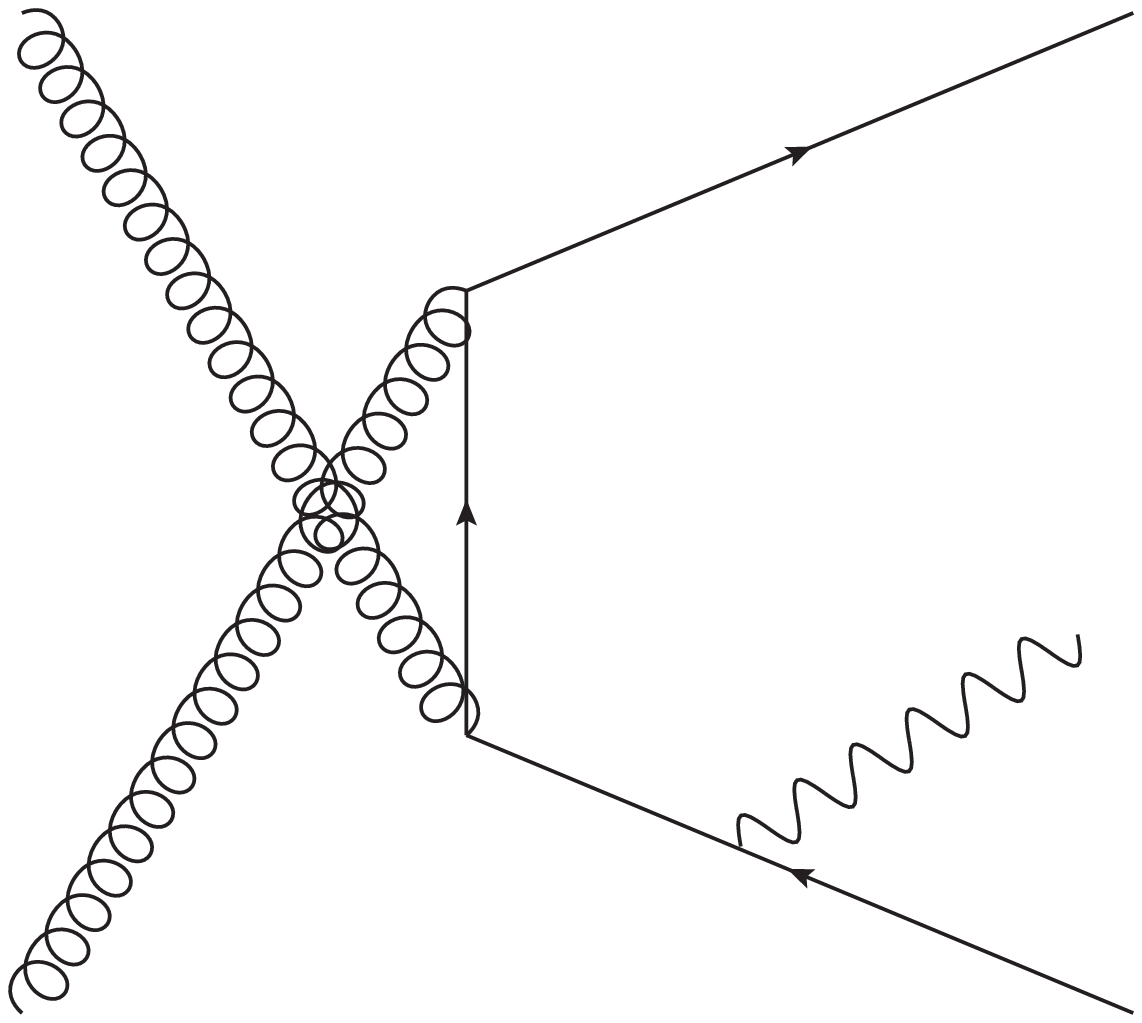} \\[1em]  
\includegraphics[width=.18\textwidth]{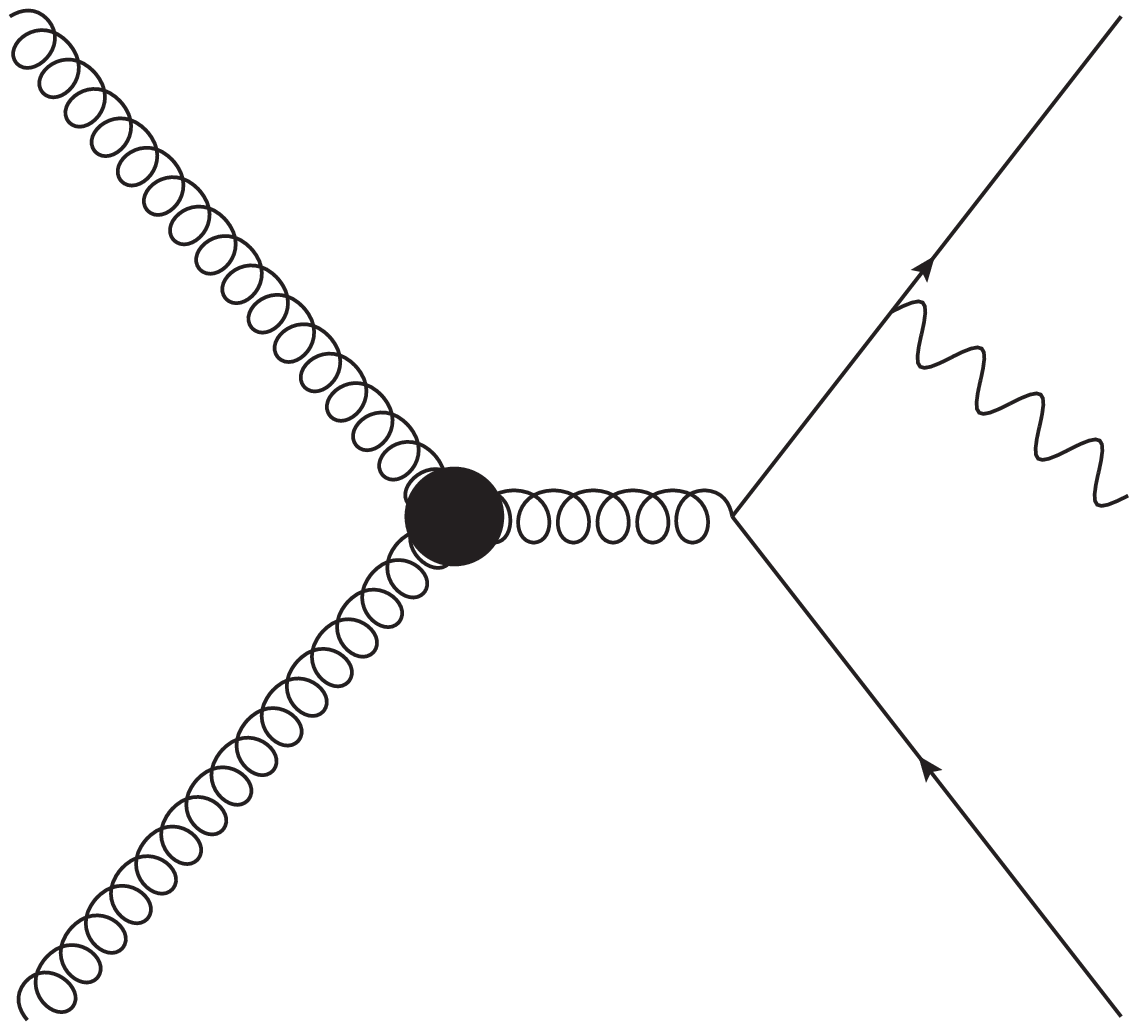} \hspace{5em} 
\includegraphics[width=.18\textwidth]{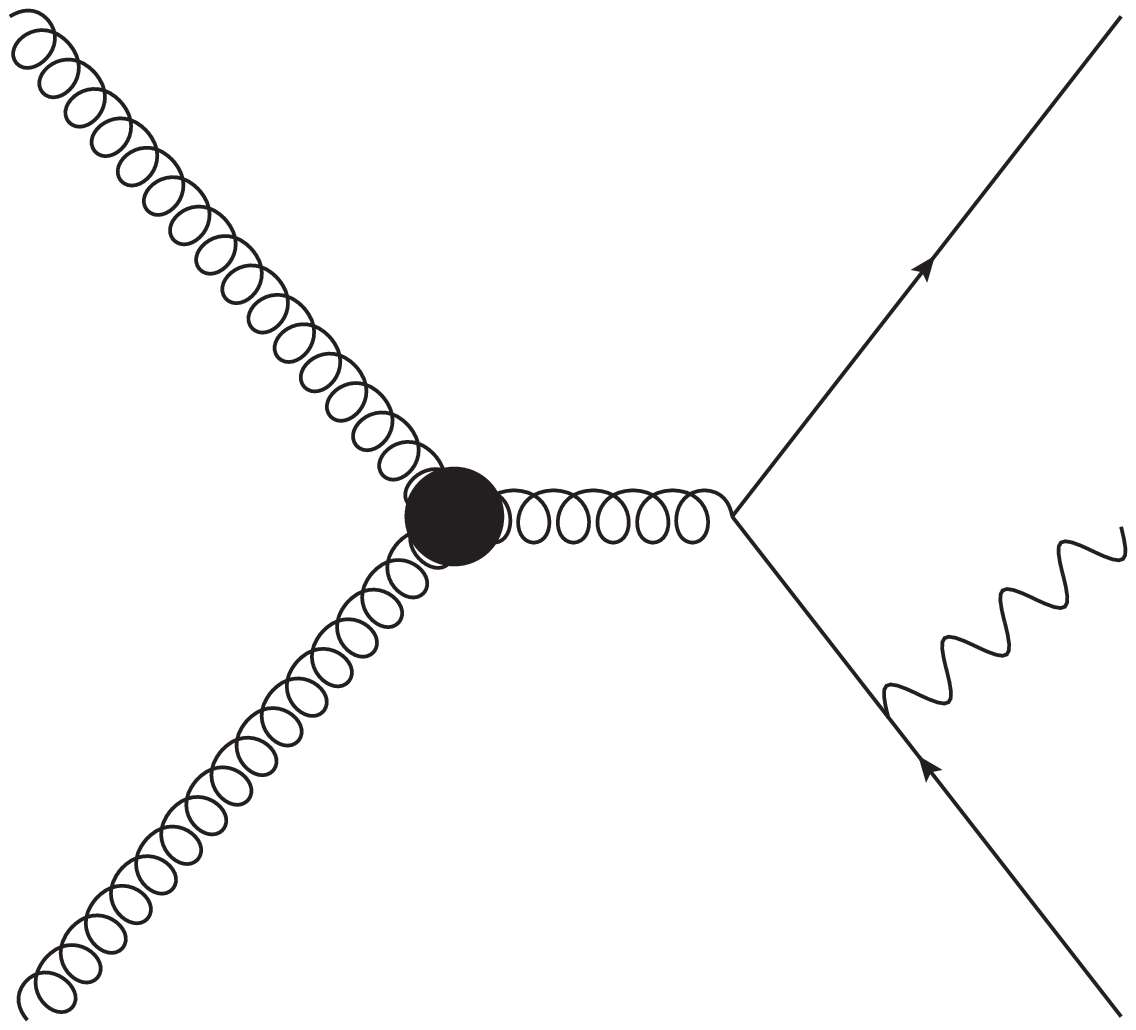} \\ 
\caption{\it Feynman diagrams that contribute to the $g^* g^* \to q\bar q V$ partonic channel corresponding to amplitudes ${\cal M}^{(i)}$ ordered in $i=1,2,\ldots,8$. The black blobs in the last two diagrams denote the effective triple gluon vertex $V_{\mathrm{eff}}$, see Appendix \ref{App:2} for the explicit definition. 
\label{Fig:ggdiags}}
\end{figure}
The $g^*(k_1) g^*(k_2) \to q(p_3)\bar q(p_4) V(q,\epsilon_{(\sigma)})\,$ hard subprocesses cross-section is calculated in the $k_T$-factorization framework. The gluons are virtual, $k_i^2 \simeq -\vec{k}_i ^2 <0$, and we shall assume the quarks to be massless, $p^2 _3 = p_4 ^4 = 0$. In the high energy limit one may decompose the gluon momenta as follows: $k_1 = x_1 P_1 + k_{1\,\perp}$ and $k_2 = x_2 P_2 + k_{2\,\perp}$. The parton level scattering amplitude is described by 8~diagrams shown in Fig.\ \ref{Fig:ggdiags}. The high energy limit for virtual gluon polarizations is used, in which the virtual gluon polarization vectors $\pi_{g^*}(k_i)$ are approximated as: 
$\pi^{\mu} _{g^*}(k_1) \simeq x_1 P^{\mu} _1 / \sqrt{\vec{k}_1 ^2} $, $\pi^{\mu} _{g^*}(k_2) \simeq x_2 P^{\mu} _2 / \sqrt{\vec{k}_2 ^2} $, the so-called ``nonsense polarizations''\footnote{The so called Collins--Ellis trick \cite{CollEll} is used in the derivations.}. Hence we introduce the impact factors ${\cal T}_{\mu} ^{(i)}$  defined as
\be
{\cal T}_{\mu} ^{(i)} = {\cal M}^{(i)} _{\mu, \alpha\beta} P_1 ^{\alpha} P_2 ^{\beta},
\ee
where ${\cal M}^{(i)} _{\mu, \alpha\beta}$ is the amplitude with amputated polarization vectors of the incoming gluons and the outgoing vector boson. The impact factors  ${\cal T}_{\mu} ^{(i)}$ corresponding to the diagrams shown in  Fig.\ \ref{Fig:ggdiags} are proportional to corresponding partonic $g^*g^*$ amplitudes ${\cal M}_{\mu} ^{(i)}$. The explicit expressions for the impact factors are given in Appendix \ref{App:2}. As was discussed in Sec.~\ref{Sec:3b}, the elementary triple gluon vertex is replaced by the effective Lipatov vertex $V_{\mathrm{eff}}$ \cite{BFKL} that may be conveniently derived e.g.\ from the Lipatov effective action \cite{EffectiveAction}. 

The $g^* g^* \to q\bar q V$ impact factor is given by ${\cal T}^{g^*g^*} _{\mu} = \sum_{i=1} ^8 {\cal T}^{(i)} _{\mu}$. This impact factor is then used to calculate $g^*g^*$ channel contributions to the Drell--Yan cross-sections $d\sigma^{(g^*g^*)} _{\sigma \sigma'}$,
\begin{eqnarray}
d\sigma^{(g^*g^*)}  _{\sigma \sigma'} = \int dx_1 \int {d^2 \vec{k}_1 \over \pi \vec{k}_1^2}  {\cal F}(x_1,\vec{k}_1^2,\mu_F) \int dx_2 \int {d^2 \vec{k}_2 \over \pi \vec{k}_2 ^2 } {\cal F}(x_2,\vec{k}_2^2,\mu_F) \hfill {} \nonumber \\
 \times \; {(2\pi)^4 {\cal H}_{\sigma \sigma'} \over 2\sp}\, dPS_3(k_1 + k_2 \to p_3 + p_4 + q), 
\end{eqnarray}
with
\be
{\cal H}_{\sigma \sigma'}  = \sum_{f}{1 \over (N_c^2 - 1)^2}\sum_{a,b}\sum_{i_3,i_4}\sum_{r_3,r_4} \,{\cal T}^{g^*g^*}_{\mu} \epsilon^{\mu} _{(\sigma)}\, \left({\cal T}^{g^*g^*}_{\nu} \epsilon^{\nu} _{(\sigma')}\right)^\dagger ,
\ee
where the summations are performed over the quark flavor $f$, the color indices $a$, $b$ of the gluons and $i_3$, $i_4$ of the quarks, and over the quark helicities $r_3$ and $r_4$. The summation over the quark helicities leads to traces over Dirac spinors which are evaluated with the FORM program for symbolic manipulations \cite{FORM}. The resulting expressions are extremely lengthy and are not explicitly displayed. They were obtained in two independent calculations and it was verified that all ${\cal H}_{\sigma \sigma'}$ tend to zero as $\vec{k}_i ^2$ when the corresponding gluon transverse momentum $\vec{k}_i ^2 \to 0$, as required by the gauge invariance condition in the high energy limit.

The phase space for the final state particles of partonic scattering is parameterized in terms of the rapidity $Y$, and the transverse momentum vector $\vec{q}_T$ of the intermediate electroweak boson, and the variables $(z,\phi_{\kappa})$ describing the $q\bar q$ kinematic configuration,   
\be
dPS_3(k_1 + k_2 \to p_3 + p_4 + q) = {dY \, d^2 q_T \, dz \, d\phi_{\kappa} \over 8(2\pi)^9}\, d\kappa^2 \, \delta \left[ \kappa^2 -   z(1-z)\left ( x_{q\bar q}x_2 S -x_{q\bar q}\frac{M_T^2 }{x_F} -  \vec{\Delta}^2\right )\right],
\ee
where the variables $z$ and $\vec \kappa$ are implicitly defined by the parameterization:
\be
p_3 = z x_{q\bar q} P_1 + {\vec{p}_3 ^2 \over z x_{q\bar q} S } P_2  +  p_{3\perp}, \qquad  
p_4 = (1-z) x_{q\bar q} P_1 + {\vec{p}_4 ^2 \over (1-z) x_{q\bar q} S } P_2  +  p_{4\perp},
\ee
and
\be
p_{3\perp} = (0,0,\vec{p}_3),\quad p_{4\perp} = (0,0,\vec{p}_4), \quad 
\vec{p}_3 = z \vec{\Delta} + \vec{\kappa}, \quad
\vec{p}_4 = (1-z) \vec{\Delta} - \vec{\kappa}, 
\ee
\be
\vec{\Delta} = \vec{k}_1 + \vec{k}_2 - \vec{q}, \quad x_{q\bar q} = x_1 - x_F.
\ee
For comparison with data it is necessary to integrate over the phase space of the final
state quark and antiquark kinematical variables, 
$
d\sigma^{(g^*g^*)}  _{\sigma\sigma'} / dY d^2 q_T = 
\int dz \int d\phi_{\kappa}  \, [d\sigma^{(g^*g^*)}  _{\sigma\sigma'} / dY d^2 q_T dz d\phi_{\kappa}]$.

\subsection{Amplitudes in the Reggeized quark--antiquark channel}
\label{Sec:3d}

For completeness, we also consider the Drell--Yan cross-sections assuming the $k_T$-factorization for quarks, see Fig.\ \ref{Fig:channels}a. Hence we consider the lowest order $q^* \bar q^*$ amplitude of virtual quark--antiquark fusion into the electroweak boson. We apply the off-shell incoming quark amplitudes and resulting hadronic cross-sections that were derived in Ref.\ \cite{Nefedov:2012cq} is so called Quark Parton Reggeization Approach. For the sea quark and antiquark TMDs ${\cal Q}_{\mathrm{sea}}$ in the proton we use the following approximation \cite{HHJ}:
\begin{eqnarray}
{\cal Q}_{\mathrm{sea}}(x, \vec{p}_T^2, \mu_F) & = & {1 \over \vec{p}_T ^2}
\int_x ^1 {dz \over z} \, \int dk_T ^2\, 
\Theta\left(\mu_F ^2 - {\vec{p}_T ^2 + z(1-z) \vec{k}_T ^2 \over  1-z}\right) \, \nonumber \\
& & \times {\alphas (\mu _F) \over 2\pi} \, P_{q^*g^*}(z,\vec{p}_T^2,\vec{k}_T ^2) \, {\cal F}(x,\vec{k}_T ^2,\mu_F),
\end{eqnarray}
where the splitting function
\be
P_{q^*g^*}(z,\vec{p}_T^2,\vec{k}_T ^2) = T_R\, \left( {\vec{p}_T ^2 \over \vec{p}_T^2 + z(1-z) \vec{k}_T ^2}\right)^2\;
\left[ (1-z)^2 + z^2 + 4z^2(1-z)^2{\vec{k}_T ^2 \over \vec{p}_T^2}\right].
\ee
This approximation assumes that the off-shell quark is produced in the last step of the $k_T$-dependent parton evolution from a splitting of the off-shell gluon. Thus the resulting $q\bar q \to V$ cross-sections represents only the sea quark--sea antiquark contribution and they approximate the $g^* g^* \to q\bar q V$ cross-sections described above. For this reason, the virtual quark--antiquark cross-sections, $d\sigma^{(q^* \bar q^*)}  _{\sigma\sigma'}$  in the Quark Parton Reggeization Approach will be used only as a reference for the more accurate $d\sigma^{(g^*g^*)}  _{\sigma\sigma'}$ cross-sections.

\section{Gluon TMD models} 
\label{Sec:4}

In this paper selected gluon TMDs are applied in the computations of the Drell--Yan cross-sections. The parameterization of the gluon TMD are the following:
\begin{itemize}

\item the Jung--Hautmann (JH) TMD, $\fjh $. This model of the gluon TMD is obtained~\cite{JHTMD} from the CCFM evolution equation \cite{CCFM}. We use the parameterization JH-2013-set1 from the TMDlib \cite{TMDlib};

\item the BFKL gluon TMD, $\fbfkl $, that emerges from the LO BFKL evolution \cite{BFKL} with parameters adjusted
to describe HERA $F_2$ data \cite{MotSadHERA} The details of the TMD are presented below;

\item a new simple model of the gluon TMD that we call the ``Weizs\"{a}cker--Williams'' (WW) gluon TMD, $\fww$ characterized by  $\propto 1/k^2_T$ behavior of the gluon TMD at large gluon transverse momenta. See the next paragraphs for the detailed description;

\item 
a quasi--collinear gluon TMD $\fgauss$ described by a narrow Gaussian distribution: $\fgauss(x,k_T^2) = {N_2}\,(1-x)^7\, \exp(-k_T^2 /  Q_{\mathrm{S}} ^2)$ inspired by the GBW gluon TMD extracted from the color dipole cross-section in the GBW saturation model, with $N_2 = 68.4$~GeV$^{-2}$, $Q_{\mathrm{S}} ^2 =\textrm{GeV}^2 ~ (x_0/x)^{\lambda}$, $x_0= 3\cdot 10^{-4}$ and $\lambda=0.29$. The very small width in $k_T$ of the gluon TMD is not realistic. This gluon TMD model is used to probe the quasi-collinear limit of the DY cross-sections useful to disentangle the parton $k_T$-effects from the effects of the emissions in the hard matrix element. 

\end{itemize}

{\bf The BFKL gluon TMD $\fbfkl (x,k_T^2)$}. It is obtained \cite{MotSadHERA} from a solution of the LO~BFKL equation assuming the input extracted from the Golec-Biernat--W\"{u}sthoff (GBW) saturation model \cite{GBW}.  Although the LO~BFKL evolution equation receives very large corrections at higher orders and the LO BFKL predictions were not expected to provide accurate description of data, the higher order effects may be  partially absorbed by redefinition of the model parameters and the LO~BFKL solution may be used as a reasonable QCD inspired model of the gluon TMD shape. In more detail, in our realization the input at gluon $x = 0.1$ for the BFKL evolution comes from the GBW model and it is very narrow in $k_T$ with a Gaussian cut-off of $k_T > 1$~GeV. At asymptotically small~$x$ the BFKL gluon TMD  scales according to the asymptotic BFKL anomalous dimension, $\fbfkl  \sim 1/k_T$. For the intermediate value of $x$, $\fbfkl $ interpolates between these two regimes and this should lead to an interesting non-trivial predictions of $\alamt$ dependence on $M$ and $\sp$.

Our model of the BFKL gluon TMD takes the following form for $x < x_{\textrm{in}}$:
\be
\fbfkl (x,k_T^2) =  {(1-x)^7 \over k_T^2} \int_{1/2 - i\infty} ^{1/2 + i \infty} {ds \over 2\pi i} (k_T^{2})^s \, \exp[\bar\alpha_{\mathrm{s}}  \chi(s) \log(x_{\textrm{in}}/x)] \tilde f_0(s).
\ee
where the LO BFKL eigenvalues are $\chi(s) = 2\psi(1) - \psi(s) - \psi (1-s)$, $\psi(z)$ is the digamma function,
\be
\tilde f_0(s) = \int_0 ^{\infty} dk_T^2 \, (k_T^2)^{-s-1}\, 
f_g(x_{\textrm{in}},k_T^2) = N_0 (Q_0^2)^{-s+1}\Gamma(2-s)
\ee
is the inverse Mellin transform of the GBW unintegrated gluon density $f_g(x,k_T^2)$ at the input  $x = x_{\textrm{in}}$: $f_g(x_{\textrm{in}},k_T^2) = N_0 k^4 \exp(-k_T^2 / Q_0^2)$. Note the difference in convention: $f_g(x,k_T^2) = k_T^2\, {\cal F}(x,k_T^2)$. In the BFKL~TMD model the exact solution of the LO~BFKL equation is multiplied by a phenomenological factor $(1-x)^7$ that has marginal influence on the gluon TMD at small~$x$ and ensures vanishing of the gluon TMD at $x\to 1$. For $x\ge x_{\textrm{in}}$, we take a phenomenological continuation to $x\to 1$:
$\fbfkl (x,k_T^2) = (1-x)^7 N_0 k_T^2 R_0^2(x)  \exp\left(-k_T^2 R_0^2(x)\right)$. Parameters of the BFKL gluon TMD model were adjusted to describe HERA $F_2$ data at small~$x$ \cite{MotSadHERA}, and they are the following: $R_0^2(x)=\frac{1}{\textrm{GeV}^2} \left( \frac{x}{9.32\cdot 10^{-4}}\right)^\lambda$, $\bar\alpha_{\mathrm{s}}  = 0.087$, $N_0 = 3.325$~GeV$^{-2}$,
$Q_0 = 0.51$~GeV and $x_{\textrm{in}} = 0.1$.

{\bf The WW gluon TMD $\fww (x,k_T^2)$. }  This distribution takes the form:
\be
\fww(x,k_T^2) = 
\begin{cases}
{(N_1 / k_0 ^2)} {(1-x)^7 \, (x^\lambda k_T^2 / k_0 ^2)^{-b}} \quad \mbox{for } \; k_T^2 \geq k_0 ^2, \\
{(N_1 / k_0 ^2)} {(1-x)^7 \, x^{-\lambda b}}  \quad \mbox{for } \; k_T^2 < k_0 ^2,
\end{cases}
\ee
where $k_0 = 1$~GeV, $\lambda = 0.29$ and $b=1$ (we introduce $b$ as a parameter in order to allow for its variations later on). The $k_T$ shape of the gluon distribution is motivated by the  $k_T$ dependence of one gluon exchange in the $t$-channel between a point-like parton and a hard probe at large momentum transfer. Such gluon exchange behaves like a virtual photon exchange, hence the corresponding virtual gluon density resembles the Weizs\"acker--Williams virtual photon density around a point-like charge. In QCD this picture of gluons as quanta emitted from point-like partons breaks down below the scale of about 1~GeV, where the color confinement effects and/or parton coherence effects in a hadron become important. 
Hence the $k_T$-dependence of $\fww(x,k_T^2)$ is frozen below $k_0 = 1$~GeV. This dynamics of gluon emission from large~$x$ partons (predominantly the valence quarks) leading to approximately $1/k_T ^2$ shape of the gluon density ${\cal F}$ was employed in the joined DGLAP / BFKL evolution equation proposed
by Kwieci\'{n}ski, Martin and Sta\'{s}to (KMS) in Ref.\ \cite{KwieMaSt}, where the non--uniform terms in the gluon TMD evolution take the $1/k_T ^2$ form (up to logarithmic modifications), and the resulting gluon TMD at moderate $x$ also scales as $1/k_T ^2$ at larger  $k_T ^2$. Unfortunately, the KMS gluon TMD is not directly applicable for gluon $x>0.01$, and cannot be used for the $Z^0$ production at the LHC. In the analysis of the data we also allow for different values of the parameter $b$ in order to probe sensitivity of $\alamt$ for the shape of the gluon.

Note that we proposed also an $x$-dependence of the WW gluon TMD. The $x$-dependence of the gluon TMD should come from the $x$-profile of the sources (predominantly the valence quarks) and from the QCD evolution. The full study of those effects is beyond the scope of this paper, so a simple model of the $x$-dependence is assumed that employs the geometric scaling property \cite{geometric} at small~$x$ where the $x$~dependence of the factor $(1-x)^7$ is mild and may be neglected. The geometric scaling parameter $\lambda$ was chosen in order to match the GBW model exponent. The factor $(1-x)^7$ is introduced to represent the gluon distribution suppression at large~$x \sim 1$, and the exponent of $1-x$ was chosen in accordance with the dimensional counting rules for spectator constituents in high energy scattering, see e.g.\ the discussion in \cite{KwMoTi}. In addition a phenomenological parameter $b$ is introduced that controls $k_T$-scaling of the WW gluon TMD model that allows to test the sensitivity of observables to the details of the TMD shape. The normalization constant $N_1 =0.889$ was adjusted by comparison to recent ATLAS data \cite{DYgamma_atlas} on the intermediate mass total Drell--Yan cross-section $d\sigma^{(\gamma^*)}  (pp\to l^+l^- X) / dM$, driven by the virtual photon exchange, see Sec.\ \ref{Sec:5}. It is important to add that the predictions for the lepton distribution angular coefficients $A_0$ and $A_2$ and for $\alamt$ weakly depend on the $x$-dependence of gluon TMDs, they are sensitive to the details of the shape in $k_T$.

A comment is in order here on the applicability of the $k_T$-factorization framework and the existing gluon TMDs in the region of gluon $x \sim 0.1$, that contributes to the large $p_T$ $Z^0$ production at the LHC. As discussed in Sec.\ \ref{Sec:3b}, the necessary approximations of the general framework do not require the limit of small gluon $x$, the approximation accuracy is controlled by the ratio $k_T / \sqrt {s_i}$, of the gluon~transverse momentum $k_T$ to the invariant mass of the gluon--target pair, $\sqrt{s_i} \simeq \sqrt{x_i S}$, that stays small enough. The existing parameterizations of the gluon TMDs, however, are not well constrained at moderate gluon~$x$. The BFKL evolution is derived assuming the small~$x$ limit, so for gluon $x \sim 0.1$ it does not provide predictions, and one probes only the assumed input distribution for the gluon TMD. The CCFM formalism is not limited to the small~$x$ domain, in particular the JH gluon TMD parameterization was extended up to $x=1$ \cite{JHTMD}. In this parameterization, however, the uncertainty of the gluon TMD grows quickly with increasing $x$ and in the region of moderate gluon $x$ is rather large. In particular, the $k_T$ shape of the gluon JH~TMD for $x \sim 0.1$ is practically unconstrained. This implies that our predictions for the DY structure functions with BFKL and JH gluon TMDs will have sizable model uncertainties. This is not, however, a fundamental feature of the approach, but rather the result of the observables chosen for fits of TMDs performed so far, e.g. in \cite{JHTMD}, the total DIS cross-section at small~$x$ was used.  
As we explicitly show in Sec.\ \ref{Sec:5} the Drell--Yan structure functions exhibit a strong sensitivity to the $k_T$-shape of the gluon TMD, therefore these observables may be used to constrain the poorly known region of moderate~$x$ and sizable $k_T$ of the gluon TMDs. A step towards this goal is the proposed WW gluon TMD, a  model of $k_T$-shape of the gluon TMD, that, as it is shown in Sec.\ \ref{Sec:5} provides an improved description of the Drell--Yan structure functions in $Z^0$ production at larger $p_T$.

\section{Results}
\label{Sec:5}

\subsection{Lam--Tung relation breaking at the $Z^0$ peak}

\noindent
\begin{figure}
\begin{tabular}{ll}
\includegraphics[width=.47\textwidth]{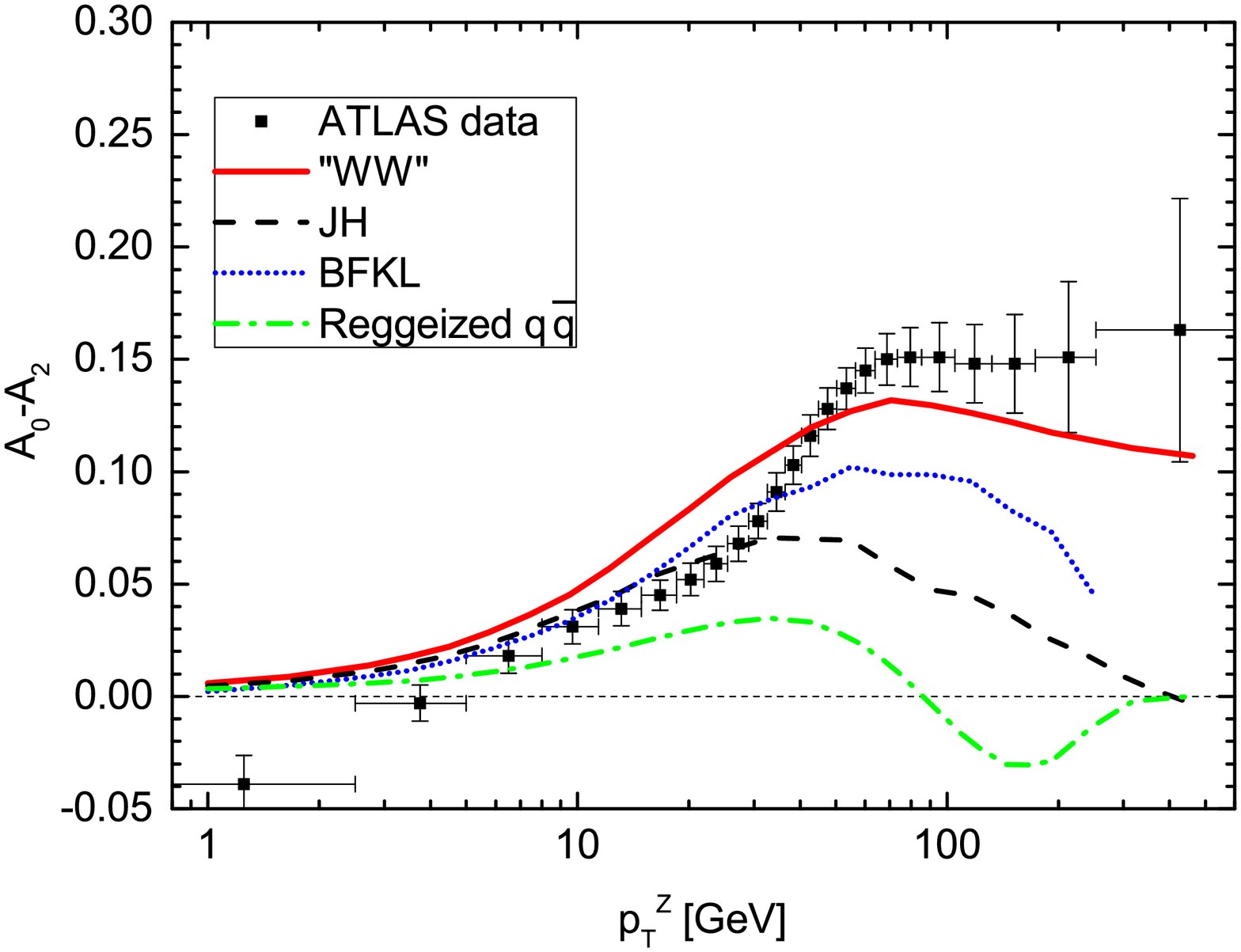} & 
\includegraphics[width=.47\textwidth]{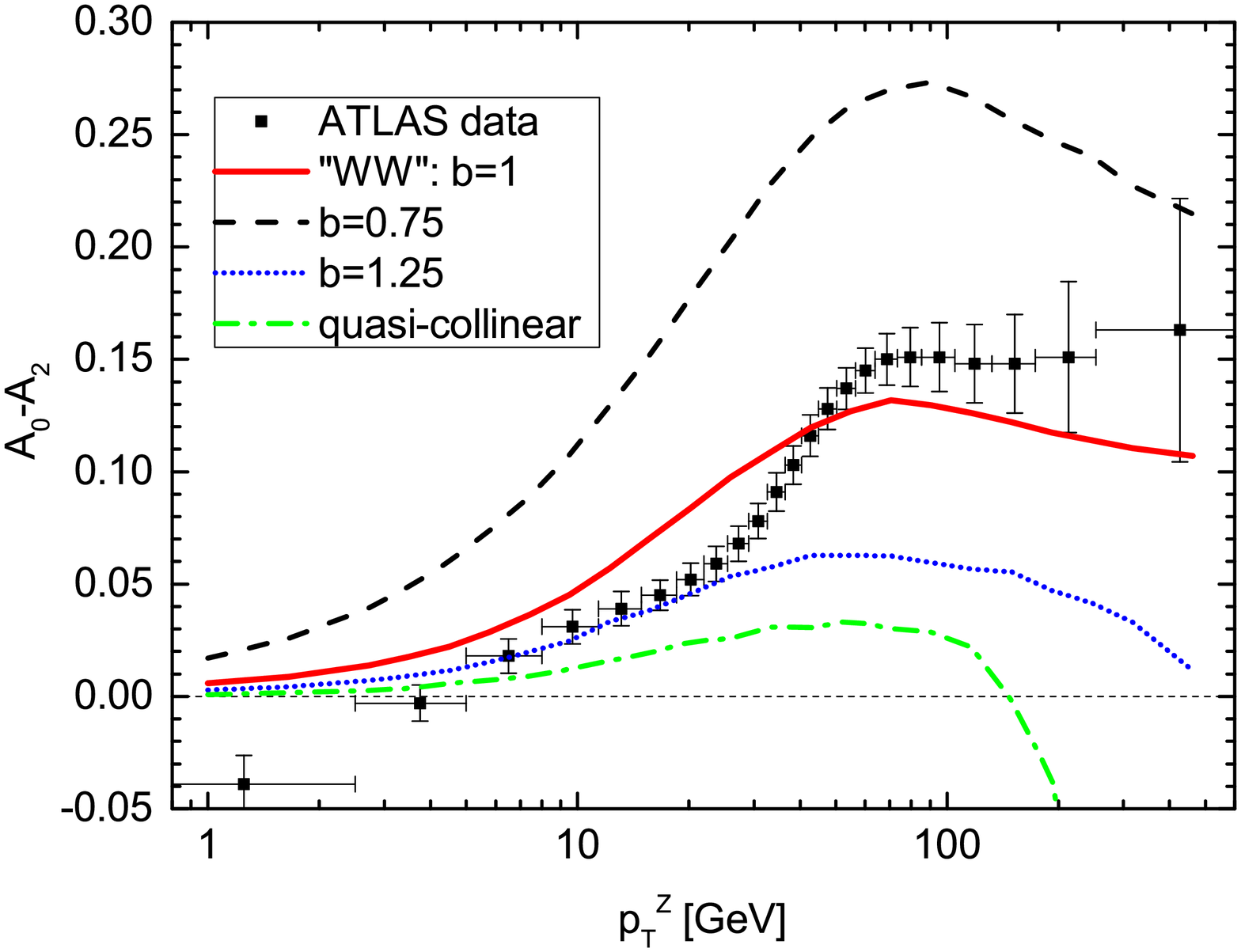} \\
{\Large a)} & {\Large b)} 
\end{tabular}
\caption{\it The Lam--Tung relation breaking coefficient $\alamt=A_0-A_2$ in the Collins--Soper frame for the $Z^0$ production in $pp$ collision at the LHC at $\sqrt{\sp} = 8$~TeV as a function of $Z^0$ transverse momentum $p_T ^Z \equiv q_T$. The ATLAS data \cite{ATLASZ0} are compared to results of calculations obtained assuming the Reggeized $q^*\bar q^*$ production channel and approximate ${\cal O}(\alphaem\alphas^2)$ $k_T$-factorization approach, assuming seven models of the gluon TMD, see the text for explanation.
\label{Fig:ALT}}
\end{figure}

Up to now the most accurate measurements of the Lam--Tung relation breaking coefficient $\alamt = A_0 - A_2$ at the LHC were performed by the ATLAS collaboration \cite{ATLASZ0} and this is the key data set for the present analysis.  Recall that up to the NNLO approximation in QCD the predictions obtained in the collinear framework do not describe the ATLAS data well \cite{ATLASZ0}. We consider contributions of both partonic channels: $\qval  g^*$ and $g^* g^*$.  The following models of the gluon TMD ${\cal F}(x,k^2,\mu_F)$ are considered: the Jung--Hautmann gluon, the BFKL gluon and the ``Weizs\"acker--Williams'' model of the gluon TMD.  For a reference we also show the predictions of the Reggeized quark model. The obtained results are compared to ATLAS data in Fig.\ \ref{Fig:ALT}a. In order to estimate effects of the gluon TMD shape in $k_T$ on $\alamt$ we also analyze different simple shape models of the gluon TMD, by altering the exponent of $1/(k^2_T)^b$ dependence of the WW gluon TMD from the central value $b=1$ to $b=0.75$ and $b=1.25$. Also we consider a quasi-collinear gluon with a narrow Gaussian profile in $k_T$, $\fgauss (x,k_T^2)$. The results of the analysis are shown in Fig.\ \ref{Fig:ALT}b. In the calculations we use NLO MMHT2014 parton distribution function parameterization \cite{MMHT} for the valence quarks and the two-loop running coupling constant $\alphas $ with $n_f = 5$ flavors and $\alphas (M_Z) = 0.12$. The mass of the $Z^0$ boson $M_Z = 91.1876$~GeV is used and in comparisons to the ATLAS data we set $\sqrt{\sp} = 8$~TeV for the observables the $Z^0$ peak. In the calculations we set the factorization scale equal to the transverse mass of the exchanged boson, $\mu_F = M_T$. The renormalisation scale is chosen to be equal to $\mu_F$.

Clearly, Fig.\ \ref{Fig:ALT} shows that the Lam--Tung breaking coefficient $\alamt$ at $Z^0$ peak is rather sensitive to the gluon TMD shape, especially at large $k_T$, where differences between used  models of ${\cal F}$ are largest. This is particularly well visible from Fig.\ \ref{Fig:ALT}b where simple analytical parameterizations of  ${\cal F}$ are tested. This means that $\alamt$ is an excellent probe of the gluon TMD. 
The quasi-collinear parameterization of the gluon TMD, which leads to much too low predictions for $\alamt$ (only about a quarter of the measured result at best), shows that even with (partial) inclusion of NNLO matrix element topologies, it is essential to take into account the incoming parton transverse momenta. A straightforward conclusion from Fig.\ \ref{Fig:ALT}b is that among the considered TMD shape models the WW gluon leads to the best overall description of data and its prediction is closest to $\alamt$ at large $Z^0$ boson transverse momentum.

\begin{figure}
\centering
\includegraphics[width=.5\textwidth]{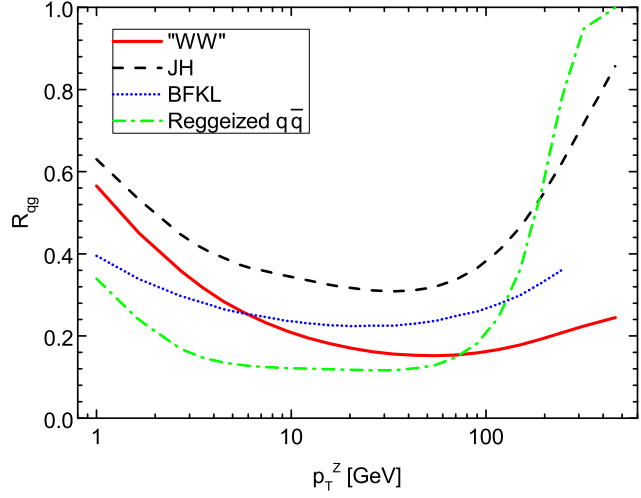}
\caption{\it Fraction $R_{qg}$ of the $qg^*$ channel contribution to the $Z^0$ cross-section integrated over the lepton angles as a function of $Z^0$ transverse momentum $p_T ^Z \equiv q_T$ in four models.  
\label{Fig:RatioqgatZ}}
\end{figure}

\noindent
\begin{figure}
\begin{tabular}{ll}
\includegraphics[width=.47\textwidth]{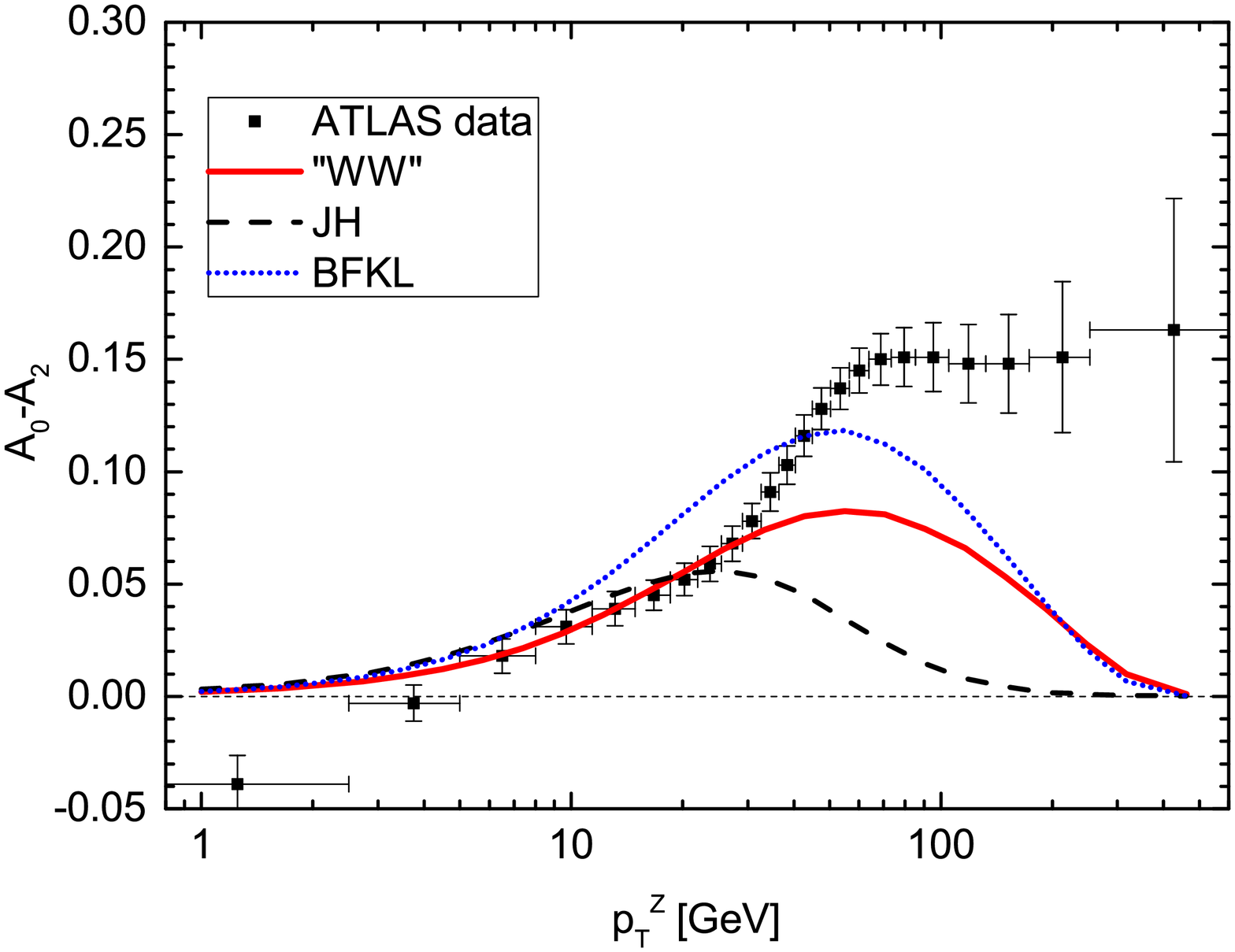} &
\includegraphics[width=.47\textwidth]{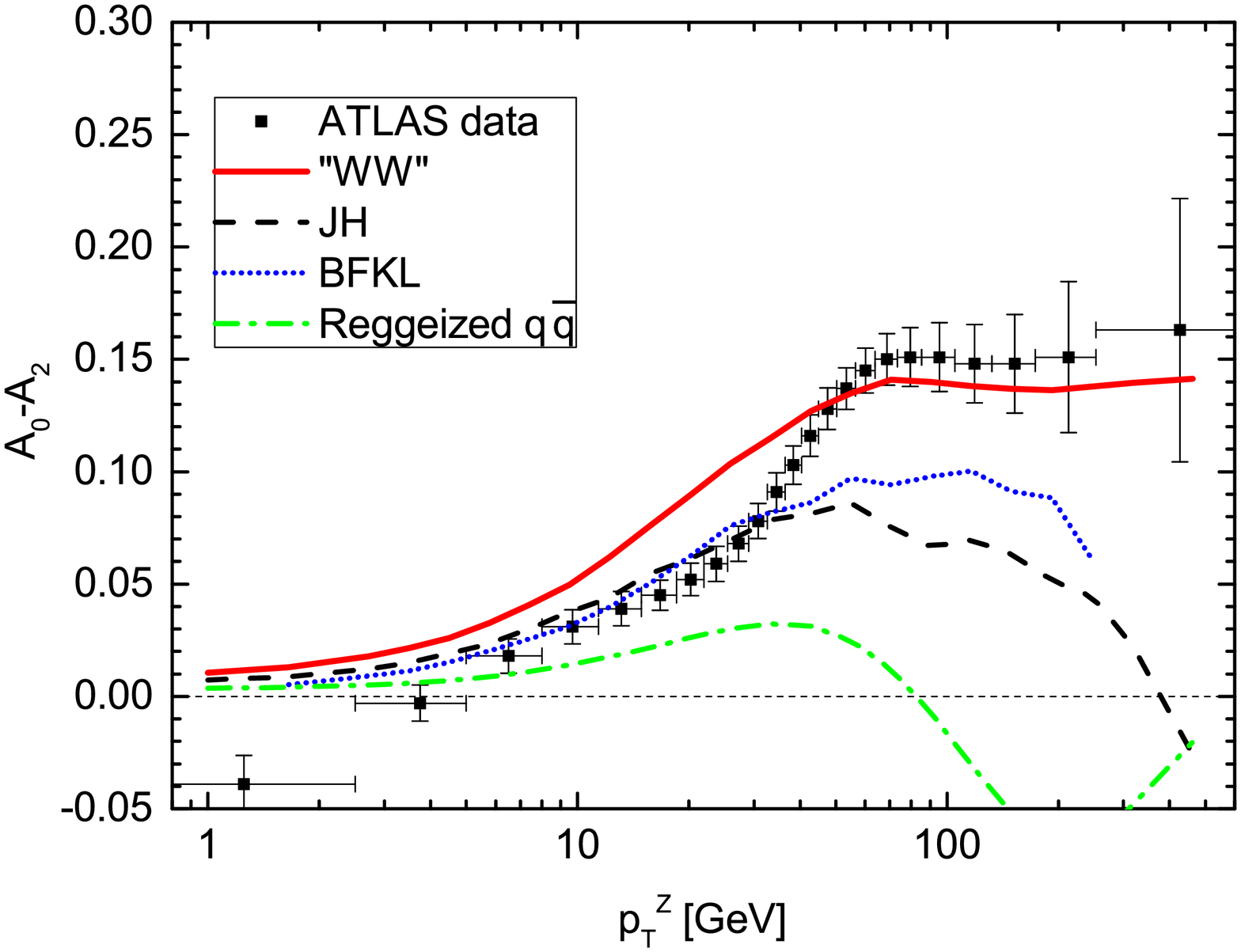}\\
{\Large a)} & {\Large b)} 
\end{tabular}
\caption{\it The Lam--Tung relation breaking coefficient $A_{LT}=A_0-A_2$ in partonic channels as a function of $Z^0$ transverse momentum $p_T ^Z \equiv q_T$: a) the $\qval  g^* \to q Z^0$ channel, and b) the $g^* g^* \to q \bar q Z^0$ channel.
\label{Fig:ALTpartonic}}
\end{figure}

In Fig.\ \ref{Fig:ALT}a predictions of the gluon TMD models are shown that are based on some underlying physics mechanisms. Hence, the Jung--Hautmann gluon TMD based on the CCFM evolution equation, the BFKL gluon TMD\footnote{Note that in Fig.\ \ref{Fig:ALT} as well and in the next figures, the BFKL TMD results are plotted up to $q_T = 200$~GeV. For larger $q_T$ larger values of the gluon TMD $x>0.1$ contribute significantly in the $g^*g^*$ channel where the BFKL~TMD is described by a very narrow distribution in $k_T$ which combined with nontrivial cancellations in the matrix elements leads to low efficiency of numerical integration and large integration uncertainties.}, the WW model inspired by the one gluon exchange $k_T$-profile, and the Reggeized quark approach with the off-shell sea quark TMD ${\cal Q}_{\mathrm{sea}}$ obtained using the JH gluon TMD.  Within this set of models the WW gluon TMD gives the best overall prescription of $\alamt$, and the BFKL model is close to the data at lower boson momenta, but falls off from the data at larger momenta. The large differences between the predicted $\alamt$ can be traced back to the difference of the gluon TMD shape at moderate $x \sim 0.1$, a typical value of gluon $x$ needed to create $q\bar q Z^0$ state at a large $Z^0$~boson transverse momentum. In this region of gluon~$x$ the JH and BFKL gluon TMDs are rather narrow in $k_T$ as a consequence of the narrow, quasi-collinear gluon input shapes and small $x$-length of the QCD evolution, still the applied model of the LL~BFKL evolution leads to higher population of the large momentum region by the gluons. The WW gluon, with its power-like behavior ${\cal F} \sim 1/k_T ^2$ at large gluon momenta for all~$x$ is much broader in $k_T$ in the relevant region of gluon~$x$, and this leads to an improved description of data. The approximation of the $g^* g^* \to q\bar q V$ matrix element by the Reggeized quark model with the Reggeized quarks coming from the virtual gluon splittings leads to the poorest description of $\alamt$.

It should be stated that the best overall description of the $\alamt$ data at $Z^0$ peak achieved with the WW model of the gluon TMD is not perfect. Our predictions overestimate the data for intermediate boson momenta and are slightly below the data at large boson momenta. Nevertheless the WW description of $\alamt$ is quite competitive with predictions of collinear QCD at the NNLO, even when combined with parton showers in the full-fledged Monte-Carlo simulation approaches \cite{ATLASZ0}. In particular, at larger boson $q_T$ our approach with the WW gluon TMD describes $\alamt$ better than full NNLO QCD predictions. Certain deviations of the description from $\alamt$ data of about 0.02--0.04 are not surprising given the simplicity of the WW gluon TMD model. We conclude from this analysis that both inclusion of gluon transverse momenta into the analysis and a ``hard'' large $k_T$ behavior of the gluon TMD at moderate~$x$ are essential for a good description of $\alamt$ at the $Z^0$ peak.

It is interesting to determine the relative importance in the $Z^0$ production of the two considered partonic channels. Hence in Fig.~\ref{Fig:RatioqgatZ} we show the ratio of the $qg^*$ channel cross-section to the sum of cross-sections from all channels,  $R_{qg} = \sigma(pp\to qg^* \to Z^0 X) /  [\sigma(pp\to g^*g^* \to Z^0 X) + \sigma(pp\to qg^* \to Z^0 X)]$, as a function of $Z^0$ transverse momentum. As seen from the figure, for the intermediate momenta the $g^*g^*$ contribution to cross-section is larger than this of $qg^*$, typically $0.2<R_{qg}<0.5$. At large boson transverse momentum, however, the $qg^*$ channel becomes increasingly important and with the JH~gluon TMD it becomes dominant. The main reason for that is the increasing values of parton~$x$ needed to produce a system with large transverse momentum, which results with the increasing importance of the valence quarks in this region, and with the decreasing width in $k_T$ of the gluon TMD. Hence it is not surprising that the estimated $g^*g^*$ channel contribution at large boson momenta is largest for the widest WW gluon and it is smallest for the most narrow JH gluon TMD. For completeness we also show the ratio $R_{qg}$ for the $q^*\bar q^*$ channel where the contribution of $g^*g^*$ channel is replaced by the one of the $q^*\bar q^*$ channel.

In the last step of the $\alamt$ analysis we show in Fig.\ \ref{Fig:ALTpartonic} the $\alamt$ obtained assuming contribution of only one of the partonic channels to the DY cross-sections. Thus, in Fig.\ \ref{Fig:ALTpartonic}a the $\alamt$ obtained  from the $qg^*$ channel and in Fig.\ \ref{Fig:ALTpartonic}b the $g^*g^*$ channel prediction is shown. The dependencies of $\alamt$ from the $qg^*$ and $g^*g^*$ channel are rather similar to each other for the BFKL gluon TMD, and with the JH gluon TMD the $g^*g^*$ channel leads to somewhat larger values of  $\alamt$ than the $qg^*$ channel. For the WW gluon TMD the $g^*g^*$ channel leads to much larger  $\alamt$ than the $qg^*$ channel. Combining the dominance of $g^*g^*$ channel contribution to the total cross-section in the WW gluon model (see Fig.\ \ref{Fig:RatioqgatZ}) and the content of Fig.\ \ref{Fig:ALTpartonic} one clearly sees that the Lam--Tung relation breaking at large boson momenta is driven by the $g^*g^*$ channel in the WW model and this feature is essential for a good description of $\alamt$  at large boson momenta. The Reggeized quark approach prediction approximating the $g^* g^*$ channel is quite far from the data. For this model the $\qval g^*$ contribution is the same as in the JH model. 

\noindent
\begin{figure}
\begin{tabular}{ll}
\includegraphics[width=.47\textwidth]{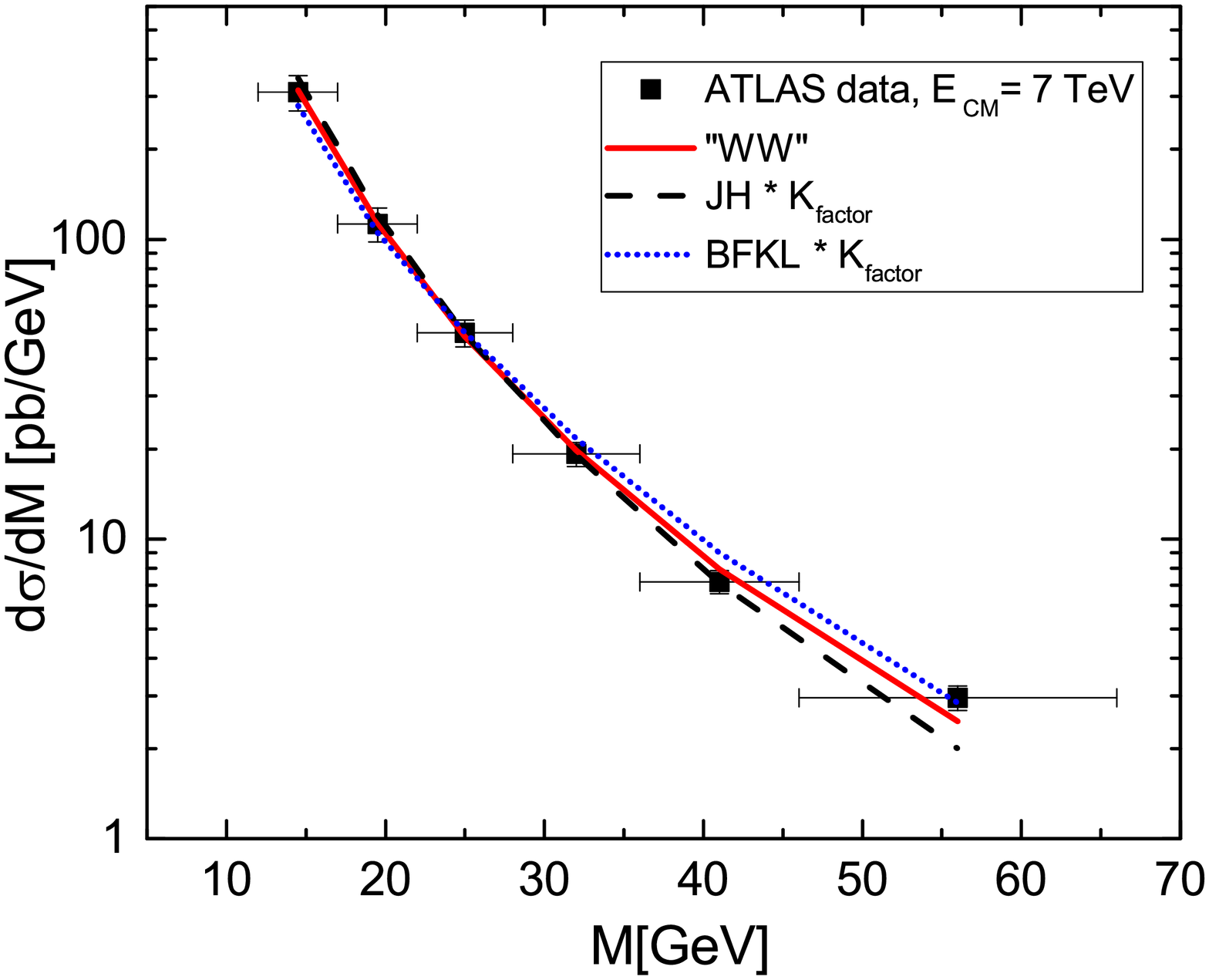} & 
\includegraphics[width=.47\textwidth]{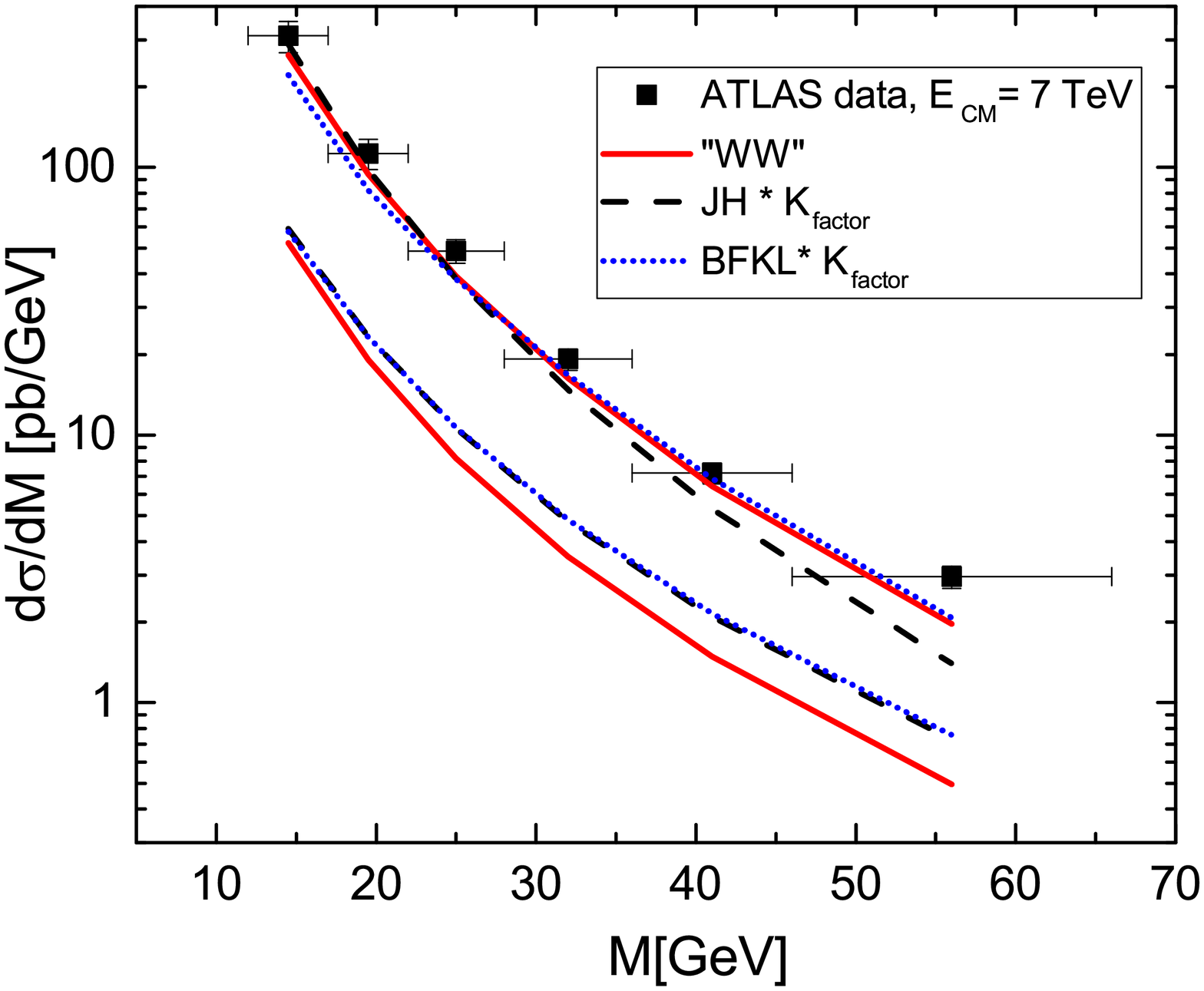}\\
{\Large a)} & {\Large b)} 
\end{tabular}
\caption{\it The cross-section for the Drell--Yan process in the low mass range in $pp$ collisions at the LHC at $\sqrt{\sp}=7$~TeV: a) the ATLAS data \cite{DYgamma_atlas} compared to predictions of three models; b) contributions of partonic channels are plotted: the $g^* g^*$ channel (upper curves) and the $\qval g^*$ channel (lower curves). 
\label{Fig:DYtot}}
\end{figure}

\subsection{Tests of the approach: $d\sigma^{(\gamma^*)}  (pp\to l^+l^- X) / dM$, $A_0$ and $A_2$}

The approach taken in the paper is based on some approximations, and our best description of the Lam--Tung relation breaking is based on a new WW model of the gluon TMD shape. Thus it is necessary to check whether the chosen approach predictions are also consistent with other Drell--Yan observables. For the checks we select the total Drell--Yan cross-section at the LHC with the intermediate virtual photon, $d\sigma^{(\gamma^*)}  (pp\to l^+l^- X) / dM$, and the coefficients $A_0$ and $A_2$ of the angular lepton distributions at the $Z^0$ peak. 

In Fig.\ \ref{Fig:DYtot}a we show descriptions of $d\sigma^{(\gamma^*)}  (pp\to l^+l^- X) / dM$ obtained with the JH, BFKL and WW gluon TMDs. In this mass region the $Z^0$ contribution is small and we neglect it. We show the ATLAS data extrapolated to the full acceptance \cite{DYgamma_atlas}. In the calculations with JH and BFKL gluon TMD a $K$-factor was applied that partially account for resummed higher order QCD corrections. The motivation to use in the DY cross-section calculation an approximate $K$-factor of the form of $K = \exp( \pi C_F \alphas (\mu_q)/2)$ with the optimal choice of the scale $\mu_q = (q_T M^2)^{1/3}$, was given in \cite{Watt:2003vf,Kulesza:1999gm} and this $K$-factor was successfully applied in several analyses of the DY process at high energies, see e.g.\ \cite{Watt:2003vf,Nefedov:2012cq,Baranov:2014ewa}. In our analysis we do not impose the full $q_T$ dependence of the $K$-factor but select a global, average value $K=1.5$ instead. For the WW model the gluon normalization was not constrained by the analysis of $\alamt$, and in order to fix it we use the data for $d\sigma^{(\gamma^*)}  (pp\to l^+l^- X) / dM$. For simplicity, the $K$-factor is absorbed here into the WW gluon normalization. A reasonably good description of the data is obtained with all three models of the gluon TMD, however the JH TMD leads to a slightly too steep $M$-dependence. The WW gluon TMD leads to the $M$-dependence consistent with the data. In Fig.\ \ref{Fig:DYtot}b, the partonic channel contributions of $g^* g^*$ and $qg^*$ are displayed separately for the three models of gluon TMD. In the considered DY pair mass range the $g^* g^*$ channel is the dominant one, with its contribution by a factor of a few greater than the contribution of the $qg^*$ channel.

\begin{figure}
{\centering
\begin{tabular}{ll}
\includegraphics[width=.4\textwidth]{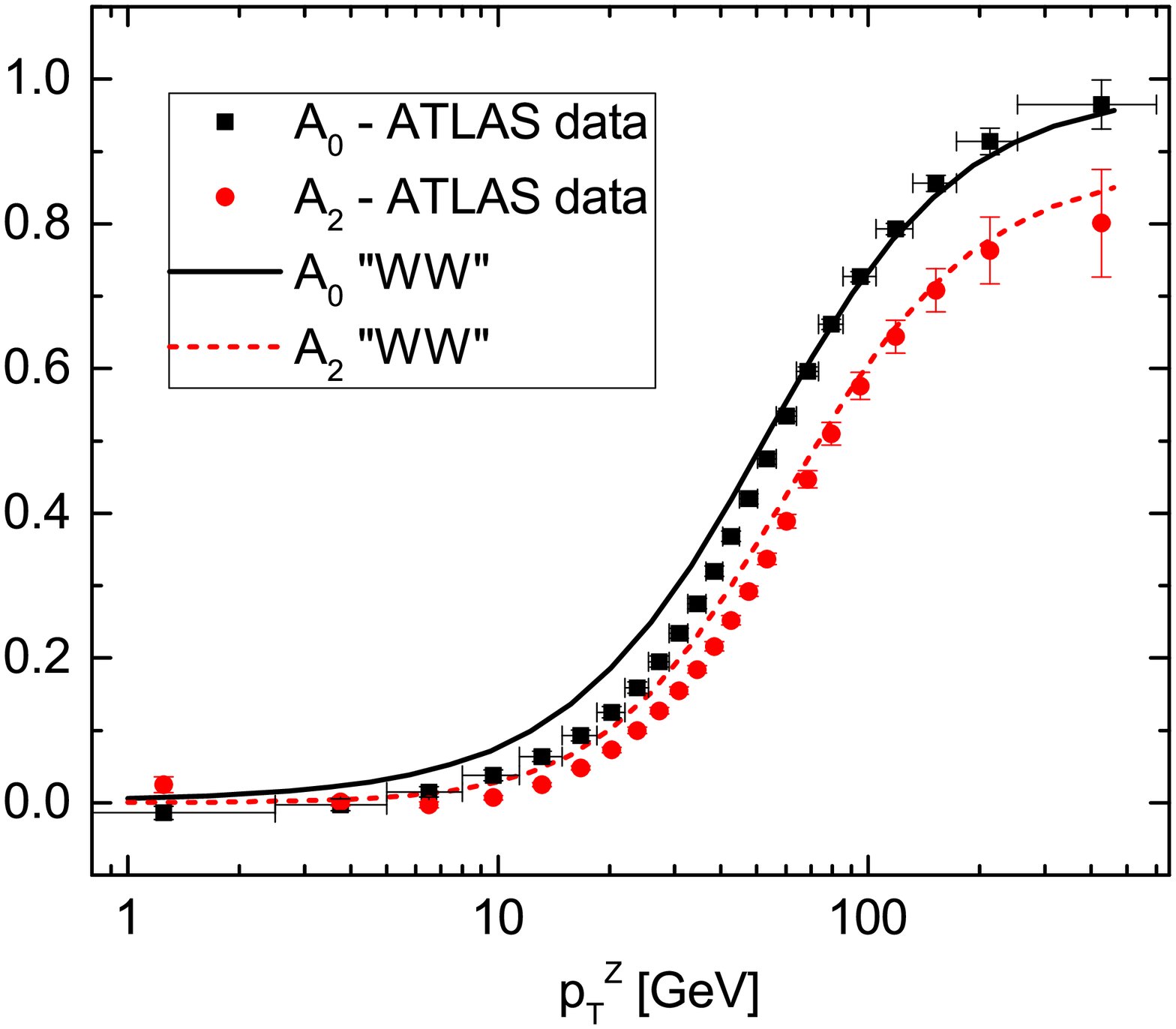}  & 
\includegraphics[width=.4\textwidth]{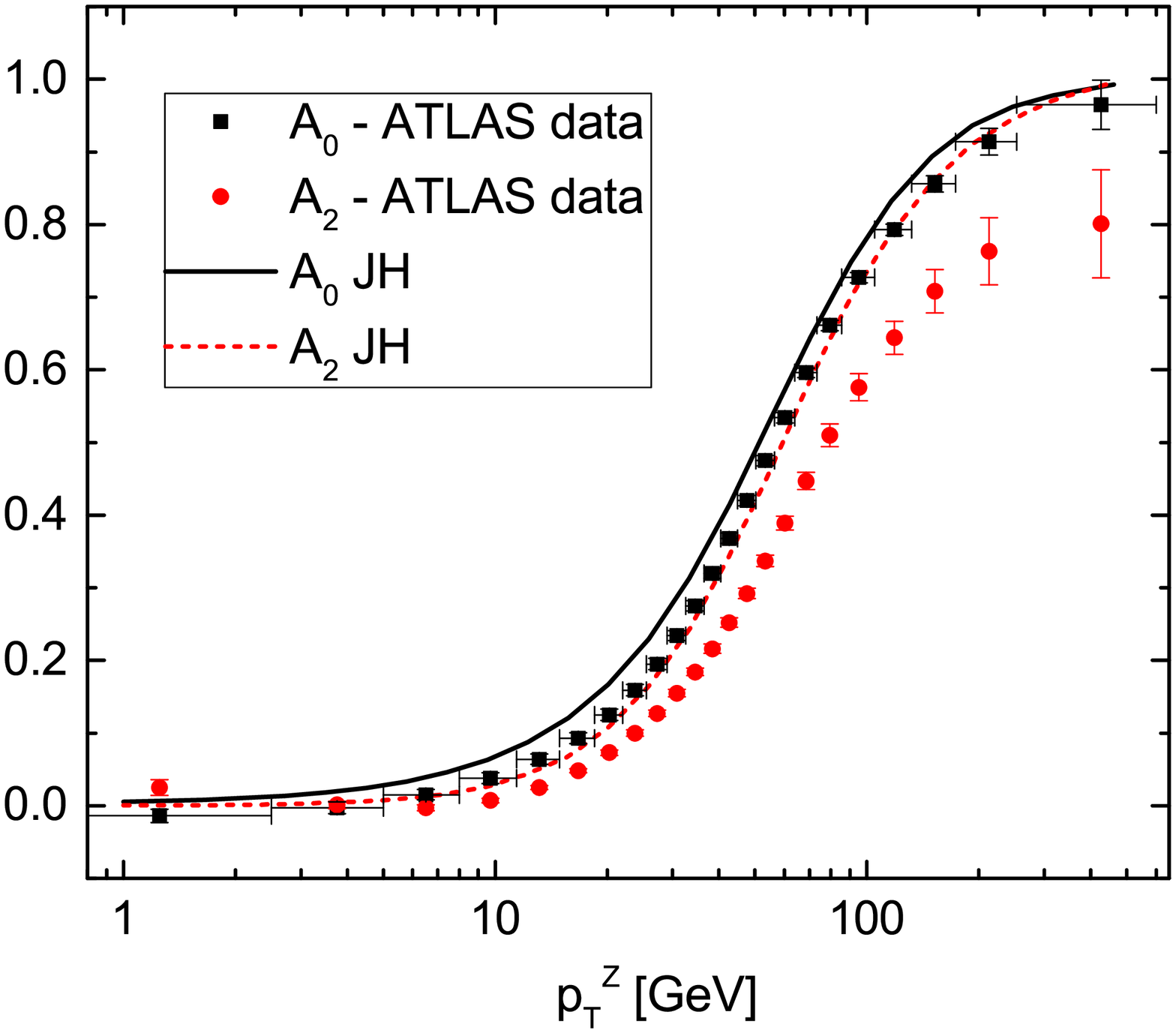} \\
{\Large a)} & {\Large b)} \vspace{3em} \\
\includegraphics[width=.4\textwidth]{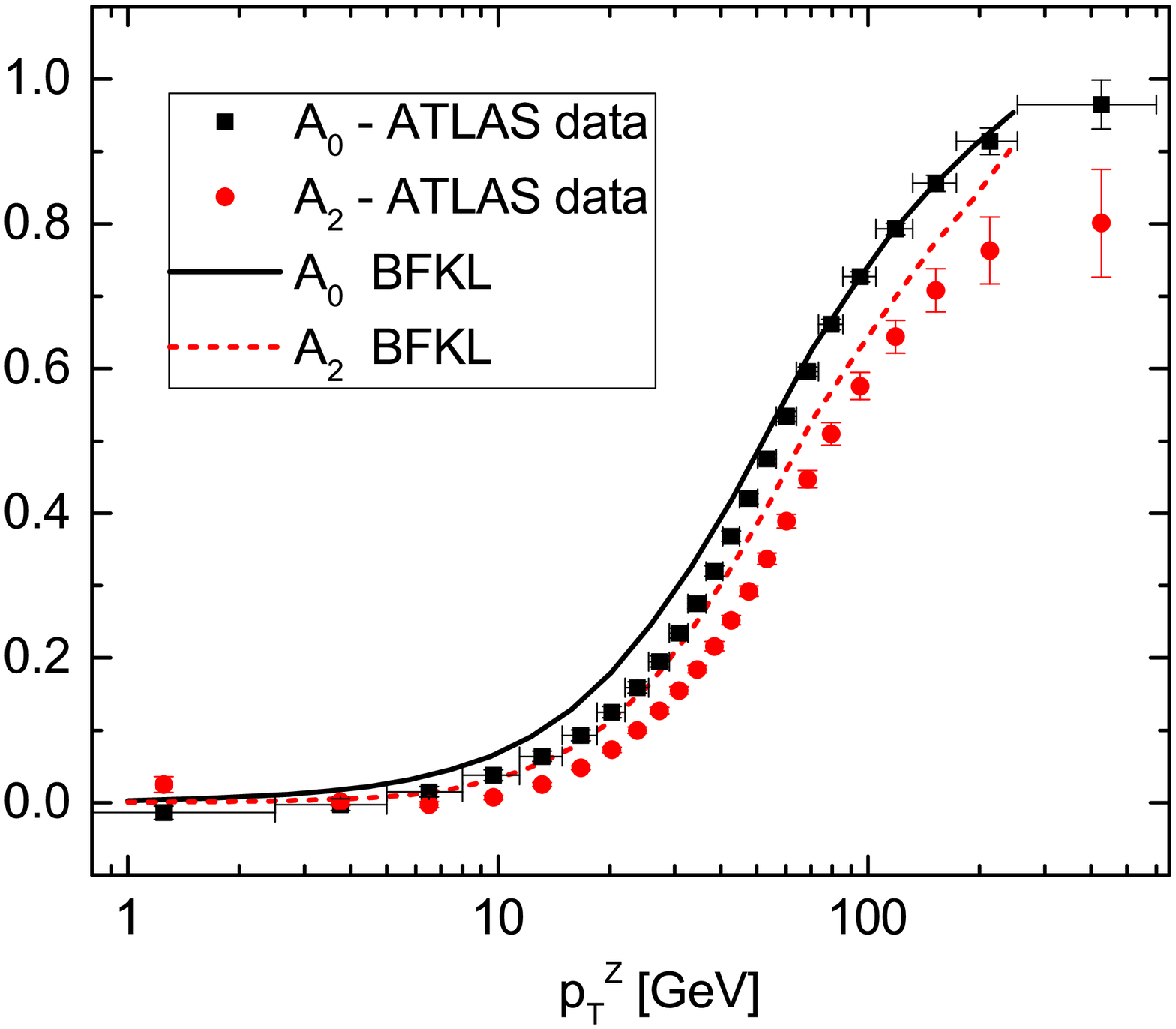} &
\includegraphics[width=.4\textwidth]{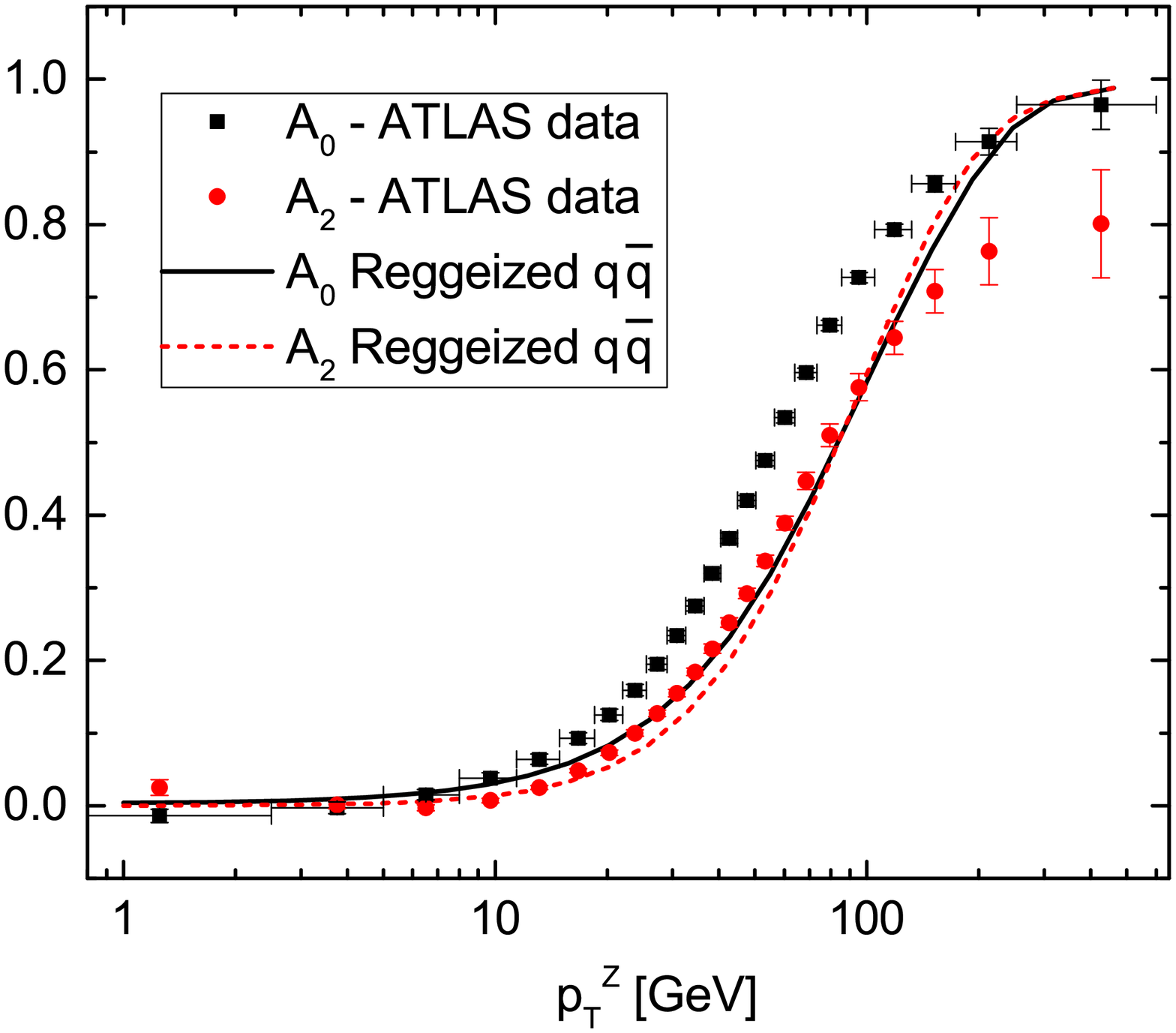} \\
{\Large c)} & {\Large d)} 
\end{tabular}\\
} 
\caption{\it Coefficients $A_0$ and $A_2$ in the Collins--Soper frame for the $Z^0$ production in $pp$ collision at the LHC at $\sqrt{\sp} = 8$~TeV as a function of $Z^0$ transverse momentum $p_T ^Z \equiv q_T$. The ATLAS data \cite{ATLASZ0} are compared to results of the approximate ${\cal O}(\alphaem\alphas^2)$ $k_T$-factorization approach, assuming models of the gluon TMD: a) the WW TMD, b) the Jung--Hautmann TMD, c) the BFKL TMD, and d) the $q^*\bar q^*$ production mechanism.
\label{Fig:A0A2}}
\end{figure}

Finally we consider the coefficients $A_0$ and $A_2$ at the $Z^0$ peak, see Fig.\ \ref{Fig:A0A2}. The ATLAS data are compared with predictions obtained with the WW gluon TMD in Fig.\ \ref{Fig:A0A2}a, JH gluon TMD in Fig.\ \ref{Fig:A0A2}b,  the BFKL gluon TMD in  Fig.\ \ref{Fig:A0A2}c,  and with the predictions of the Reggeized quark model  in  Fig.\ \ref{Fig:A0A2}d. All the models predictions have similar shapes to the data but they do not describe the data within experimental errors. Among these models the best simultaneous description of $A_0$ and $A_2$ is obtained with the WW~gluon TMD. This model results agree with the data for both $A_0$ and $A_2$ within errors except of a region of intermediate $Z^0$ boson momenta, between 5~GeV and 50~GeV, where the WW model slightly overestimates the data for both coefficients. 

To summarize, from the simultaneous analysis of $\alamt$, $A_0$ and $A_2$ at the $Z^0$ peak and $d\sigma^{(\gamma^*)} (pp\to l^+l^- X)  / dM$ one concludes that the proposed WW model of the gluon TMD describes reasonably well all these observables, and provides a competitive or the best description among the studied models of the data over most of the kinematical range.

\section{Discussion}

The main conclusions coming from the analysis of the Lam--Tung relation breaking at the $Z^0$ peak within an
approximate ${\cal O}(\alphaem\alphas^2)$ $k_T$-factorization approach are the following: (i) the gluon off-shellness effects have an essential impact on $\alamt$ up to the NNLO in QCD; (ii) the angular distributions of Drell--Yan dileptons provide a sensitive probe of the gluon TMD with high discrimination power; (iii) for a correct description of the Lam--Tung relation breaking at large boson momenta it is essential to use a gluon TMD with a hard $\sim 1/k_T^2$ behavior at large $k_T$. 

The validity of these conclusions relies upon the important assumption of our approach that the performed selection of the NNLO contributions does not affect the results significantly. In fact, on the top of the applied high-energy approximation, the terms of $\qval g^* \to qgV$ and one loop corrections to $\qval g^* \to qV$ are not taken into account. These terms are not treated in the present paper as the one loop computations in $k_T$-factorization framework still pose a serious theoretical challenge that deserves a separate study. It may be argued, however, that the neglected terms should not alter our main conclusions. The key argument comes from the dominance of the $g^*g^* \to q\bar q V$ channel over the $\qval  g^* \to q\bar V$ channel in the total Drell--Yan cross-section at $Z^0$ peak over most of the boson transverse momentum range, see Fig.\ \ref{Fig:RatioqgatZ}. Next, the one loop corrections to  $\qval g^* \to qV$ and the contribution of the NNLO channel $\qval g^* \to qgV$ combine together into the ${\cal O}(\alphas )$ correction to the  $\qval  g^* \to q V$ channel (this combination is necessary to be performed, as it is needed for the cancellation of infra-red divergences). So, the two neglected contributions of the NNLO enter only as a ${\cal O}(\alphas )$ correction to a sub-dominant partonic channel. Therefore neglecting these terms should have only a limited impact on the predicted observables and should not affect the main features of the result. Moreover the approximation applied should work with even better accuracy for smaller DY dilepton masses and/or for larger beam collision energy  $\sqrt{\sp}$. Still it is a natural and desired direction of further analysis to complete the calculation of NNLO Drell--Yan cross-section in the $k_T$-factorization framework.  

The description of $\alamt$ obtained in this paper is not perfect yet and the application of a phenomenological model of the gluon TMD for the best description is not fully satisfactory. The next steps in this direction should be to combine the extracted information on the relevance of the Weizs\"acker--Williams picture of the gluon TMD emerging as consequence of the gluon emissions from the point-like partons at larger~$x$, with a QCD evolution equations for the off-shell gluon density. This may be done e.g.\ by an extension of the KMS formalism to larger~$x$ or by suitably adjusting the input and non-uniform terms in the gluon TMD evolution.

It is interesting to note that the approximation of the $g^* g^* \to q\bar q V$ matrix element by a hard matrix element with virtual quarks $q^* \bar q ^* \to V$ preceded by the splitting of the off-shell gluons, $g^* \to q^*$ and $g^* \to \bar q^*$, does not work well in theoretical estimates of $A_0$, $A_2$ and $\alamt$. It should be interesting to identify the reason why this approximation is not accurate in this application.  

Taking into account variety and precision of already performed and coming measurements of DY lepton observables and their sensitivity to the gluon TMD features, DY process should serve as the key data set for constraining parton transverse momentum distributions.

\section*{Acknowledgements}
We are indebted to E.\ Richter-W\!\k{a}s for many interesting
discussions. We thank H.~Jung, F.~Hautmann and J.\ Bartels for their
interest in the results and useful comments.
Support of the Polish National Science Centre grants no.\
DEC-2014/13/B/ST2/02486 is gratefully acknowledged.
TS acknowledges support in the form of a scholarship of Marian
Smoluchowski Research Consortium Matter Energy Future from KNOW funding.


\appendix
\section*{Appendix}

\section{Transformation from the Gottfried--Jackson helicity frame to the Collins--Soper frame}
\label{App:1}

The transformation of the Drell--Yan cross-sections in the helicity basis from the Gottfried--Jackson \cite{Gottfried:1964nx} helicity frame to the Collins--Soper frame \cite{Collins_Soper_frame} is described by a linear transformation. In Sec.\ \ref{Sec:3c} this transformation is denoted by ${\cal R}_{\tau\tau'}$. The explicit form of the transformation is the following:
\begin{eqnarray}
\Hi_L    & = & {M^2 \tHi_L   + 2M q_T \tHi_{LT} + \left(\tHi _T - \tHi _{TT}\right) q_T ^2 \over M_T ^2}, \nonumber \\
\Hi_T    & = & {M^2 \tHi_T - M q_T \tHi_{LT} + \left(\tHi _L + \tHi_T + \tHi _{TT}\right) q_T ^2 / 2 \over M_T ^2}, \nonumber \\
\Hi_{TT}  & = & {M^2 \tHi_{TT} + M q_T \tHi_{LT} + \left(-\tHi _L + \tHi_T + \tHi _{TT}\right) q_T ^2 / 2 \over M_T ^2}, \nonumber \\
\Hi_{LT}  & = & {(q_T^2-M^2) \tHi_{LT} + M q_T \left(\tHi_L- \tHi_T + \tHi_{TT}\right)  \over M_T ^2}. \nonumber \\
\end{eqnarray} 

\section{Impact factors of diagrams contributing to the $g^*g^* \to q\bar q V$ process}
\label{App:2}

The impact factors corresponding to the diagrams shown in  Fig.\ \ref{Fig:ggdiags} are the following,
\begin{eqnarray}
{\cal T}_{\mu} ^{(1)} & = & e e_{f} g_s^2  \; \bar u^{i_3} _{r_3} (p_3) (t^a t^b) \left[\gamma_{\mu}{\hat{v}_1 \over v_1^2}\hat{P}_1{\hat{v}_2 \over v_2 ^2}\hat{P}_2\right]  v^{i_4} _{r_4} (p_4), \nonumber\\
{\cal T}_{\mu} ^{(2)} & = & e e_{f} g_s^2 \; \bar u^{i_3} _{r_3} (p_3) (t^a t^b) \left[\hat P_1 {\hat{v}_3 \over v_3^2}\gamma_{\mu}{\hat{v}_2 \over v_2 ^2}\hat{P}_2\right]  v^{i_4} _{r_4} (p_4), 
\nonumber \\
{\cal T}_{\mu} ^{(3)} & = &  e e_{f} g_s^2 \; \bar u^{i_3} _{r_3} (p_3) (t^a t^b) \left[\hat P _1 {\hat{v}_3 \over v_3^2}\hat{P}_2{\hat{v}_4 \over v_4 ^2}\gamma_{\mu}\right]  v^{i_4} _{r_4} (p_4), 
\nonumber \\
{\cal T}_{\mu} ^{(4)} & = &  e e_{f} g_s^2 \;  \bar u^{i_3} _{r_3} (p_3) (t^b t^a) \left[\gamma_{\mu}{\hat{v}_1 \over v_1^2}\hat{P}_2{\hat{v}_5 \over v_5 ^2}\hat{P}_1\right]  v^{i_4} _{r_4} (p_4), 
\nonumber \\
{\cal T}_{\mu} ^{(5)} & = &  e e_{f} g_s^2 \; \bar u^{i_3} _{r_3} (p_3) (t^b t^a) \left[\hat P_2{\hat{v}_6 \over v_6^2}\gamma_{\mu}{\hat{v}_5 \over v_5 ^2}\hat{P}_1\right]  v^{i_4} _{r_4} (p_4), 
\nonumber \\
{\cal T}_{\mu} ^{(6)} & = &  e e_{f} g_s^2 \; \bar u^{i_3} _{r_3} (p_3) (t^b t^a) \left[\hat P_2 {\hat{v}_6 \over v_6^2}\hat{P}_1{\hat{v}_4 \over v_4 ^2}\gamma_{\mu}\right]  v^{i_4} _{r_4} (p_4), 
\nonumber \\
{\cal T}_{\mu} ^{(7)} & = & -i e e_{f} g_s^2 \; { V^{\alpha} _{\mathrm{eff}}  \over (k_1 + k_2)^2 }  \, f^{abc} \,  
\bar u^{i_3} _{r_3} (p_3) t^c \left[ \gamma_{\mu}  {\hat{v}_1 \over v_1 ^2} \gamma_{\alpha} \right] v^{i_4} _{r_4} (p_4), \nonumber \\ 
{\cal T}_{\mu} ^{(8)} & = &  -i e e_{f} g_s^2 \; {V^{\alpha} _{\mathrm{eff}}   \over (k_1 + k_2)^2 } \, f^{abc} \,
\bar u^{i_3} _{r_3} (p_3) t^c \left[ \gamma_{\alpha}  {\hat{v}_4 \over v_4 ^2} \gamma_{\mu} \right] v^{i_4} _{r_4} (p_4), 
\end{eqnarray}
where the three gluon effective vertex is given by \cite{EffectiveAction}:
\begin{equation}
V _{\mathrm{eff}} ^{\alpha} = {S\over 2} (k_{2} ^{\alpha} - k_{1} ^{\alpha}) + 
  \left( 2 P_2\cdot k_1  +  {P_1 \cdot P_2 \over P_1 \cdot k_2} k^2 _1 \right) P_{1} ^{\alpha}
- \left( 2 P_1\cdot k_2  +  {P_1 \cdot P_2 \over P_2 \cdot k_1} k^2 _2 \right) P_{2} ^{\alpha},
\end{equation}
the four momenta carried by the intermediate lines are: 
\be
v_1 = p_3 + q, \qquad v_2 = p_3 + q - k_1,\qquad v_3 = p_3 - k_1, 
\ee
\be
v_4 = p_3 - k_1 - k_2 , \qquad v_5 = p_3 + q - k_2,\qquad v_6 = p_3 - k_2,
\ee
and $t^a$~are Gell--Mann matrices. The quark spinors $u^{i_3} _{r_3}$ and $v^{i_4} _{r_4}$ carry helicities $r_3$ and $r_4$, and colors $i_3$ and $i_4$. Gluon colors are denoted by $a$ and $b$ for the external gluons
and $c$ for the exchanged gluon. Note that we consistently assume the zero mass for the quarks. In the above formulae we adopt the notation $\hat v = v^{\mu} \gamma_{\mu}$. Our result agrees with expressions from Refs. \cite{DeakSchwen} and \cite{BLZdy}.


\begin{thebibliography}{99}

\bibitem{DrellYan}
  S.~D.~Drell and T.~M.~Yan,
  Phys.\ Rev.\ Lett.\  {\bf 25} (1970) 316;
   Erratum: [Phys.\ Rev.\ Lett.\  {\bf 25} (1970) 902].


\bibitem{LT12}
  C.~S.~Lam and W.~K.~Tung,
  Phys.\ Rev.\ D {\bf 18} (1978) 2447;
  Phys.\ Rev.\ D {\bf 21} (1980) 2712.



\bibitem{DYgamma_atlas} 
  G.~Aad {\it et al.} [ATLAS Collaboration],
  JHEP {\bf 1406} (2014) 112;
[arXiv:1404.1212 [hep-ex]].


\bibitem{CMS:2014jea}
  V.~Khachatryan {\it et al.} [CMS Collaboration],
  Eur.\ Phys.\ J.\ C {\bf 75} (2015) 147;
 [arXiv:1412.1115 [hep-ex]].


\bibitem{Aad:2014xaa}
  G.~Aad {\it et al.} [ATLAS Collaboration],
JHEP {\bf 1409} (2014) 145;
[arXiv:1406.3660 [hep-ex]].
 


\bibitem{Khachatryan:2015oaa}
  V.~Khachatryan {\it et al.} [CMS Collaboration],
  Phys.\ Lett.\ B {\bf 749} (2015) 187;
[arXiv:1504.03511 [hep-ex]].



\bibitem{CMSLT}
  V.~Khachatryan {\it et al.} [CMS Collaboration],
  Phys.\ Lett.\ B {\bf 750} (2015) 154;
[arXiv:1504.03512 [hep-ex]].


\bibitem{ATLASZ0}
  G.~Aad {\it et al.} [ATLAS Collaboration],
  JHEP {\bf 1608} (2016) 159;
 [arXiv:1606.00689 [hep-ex]].


\bibitem{DYNNLO}
  E.~Mirkes and J.~Ohnemus,
  Phys.\ Rev.\ D {\bf 51} (1995) 4891;
[hep-ph/9412289].


\bibitem{Karlberg:2014qua}
  A.~Karlberg, E.~Re and G.~Zanderighi,
  JHEP {\bf 1409} (2014) 134;
[arXiv:1407.2940 [hep-ph]].





\bibitem{Berger:1979du}
  E.~L.~Berger and S.~J.~Brodsky,
  Phys.\ Rev.\ Lett.\  {\bf 42} (1979) 940.

\bibitem{Brandenburg:1994wf}
  A.~Brandenburg, S.~J.~Brodsky, V.~V.~Khoze and D.~Mueller,
  Phys.\ Rev.\ Lett.\  {\bf 73} (1994) 939;
[hep-ph/9403361].

\bibitem{Eskola:1994py}
  K.~J.~Eskola, P.~Hoyer, M.~Vanttinen and R.~Vogt,
  Phys.\ Lett.\ B {\bf 333} (1994) 526;
[hep-ph/9404322].


\bibitem{MSS}
  L.~Motyka, M.~Sadzikowski and T.~Stebel,
  JHEP {\bf 1505} (2015) 087;
[arXiv:1412.4675 [hep-ph]].

\bibitem{Brzeminski:2016lwh}
  D.~Brzeminski, L.~Motyka, M.~Sadzikowski and T.~Stebel,
  JHEP {\bf 1701} (2017) 005;
  [arXiv:1611.04449 [hep-ph]].

 

\bibitem{Boer:1999mm}
  D.~Boer,
  Phys.\ Rev.\ D {\bf 60} (1999) 014012;
[hep-ph/9902255].

\bibitem{Boer:2006eq}
  D.~Boer and W.~Vogelsang,
  Phys.\ Rev.\ D {\bf 74} (2006) 014004;
[hep-ph/0604177].

\bibitem{Berger:2007jw}
  E.~L.~Berger, J.~W.~Qiu and R.~A.~Rodriguez-Pedraza,
  Phys.\ Rev.\ D {\bf 76} (2007) 074006;
[arXiv:0708.0578 [hep-ph]].


\bibitem{Peng:2015spa}
  J.~C.~Peng, W.~C.~Chang, R.~E.~McClellan and O.~Teryaev,
  Phys.\ Lett.\ B {\bf 758} (2016) 384;
[arXiv:1511.08932 [hep-ph]].






\bibitem{GLR}
  L.~V.~Gribov, E.~M.~Levin and M.~G.~Ryskin,
  Phys.\ Rept.\  {\bf 100} (1983) 1.
 


\bibitem{BFKL}
  E.~A.~Kuraev, L.~N.~Lipatov and V.~S.~Fadin,
  Sov.\ Phys.\ JETP {\bf 45} (1977) 199
   [Zh.\ Eksp.\ Teor.\ Fiz.\  {\bf 72} (1977) 377];
  I.~I.~Balitsky and L.~N.~Lipatov,
  Sov.\ J.\ Nucl.\ Phys.\  {\bf 28} (1978) 822
   [Yad.\ Fiz.\  {\bf 28} (1978) 1597].
 

\bibitem{RWW}
  E.~Richter-W\!\k{a}s and Z.~W\!\k{a}s,
  Eur.\ Phys.\ J.\ C {\bf 76} (2016) 473;
[arXiv:1605.05450 [hep-ph]];
%
  arXiv:1609.02536 [hep-ph].



\bibitem{bfklrev}
  L.~N.~Lipatov,
  Phys.\ Rept.\  {\bf 286} (1997) 131;
  [hep-ph/9610276].



\bibitem{CCFM}  M.~Ciafaloni,
  Nucl.\ Phys.\ B {\bf 296} (1988) 49;
%
  S.~Catani, F.~Fiorani and G.~Marchesini,
  Phys.\ Lett.\ B {\bf 234} (1990) 339;
%
  G.~Marchesini,
  Nucl.\ Phys.\ B {\bf 445} (1995) 49;
[hep-ph/9412327].




\bibitem{CaCiaHaut}
  S.~Catani, M.~Ciafaloni and F.~Hautmann,
  Phys.\ Lett.\ B {\bf 242} (1990) 97;
%
  Nucl.\ Phys.\ B {\bf 366} (1991) 135.




\bibitem{TMDrev}
  R.~Angeles-Martinez {\it et al.},
  Acta Phys.\ Polon.\ B {\bf 46} (2015) 2501;
 [arXiv:1507.05267 [hep-ph]].



\bibitem{QCDdipole}
  N.~N.~Nikolaev and B.~G.~Zakharov,
  Z.\ Phys.\ C {\bf 49} (1991) 607.
   





\bibitem{Brodsky}
  S.~J.~Brodsky, A.~Hebecker and E.~Quack,
  Phys.\ Rev.\ D {\bf 55} (1997) 2584;
 [hep-ph/9609384].


\bibitem{Kopeliovich}
  B.~Z.~Kopeliovich, J.~Raufeisen and A.~V.~Tarasov,
  Phys.\ Lett.\ B {\bf 503} (2001) 91;
 [hep-ph/0012035];
%
  B.~Z.~Kopeliovich, J.~Raufeisen, A.~V.~Tarasov and M.~B.~Johnson,
  Phys.\ Rev.\ C {\bf 67} (2003) 014903;
  [hep-ph/0110221].


\bibitem{GJ}
  F.~Gelis and J.~Jalilian-Marian,
  Phys.\ Rev.\ D {\bf 66} (2002) 094014;
  [hep-ph/0208141];
  Phys.\ Rev.\ D {\bf 76} (2007) 074015;
  [hep-ph/0609066].


\bibitem{GolecBiernat:2010de}
  K.~Golec-Biernat, E.~Lewandowska and A.~M.~Sta\'{s}to,
  Phys.\ Rev.\ D {\bf 82} (2010) 094010;
 [arXiv:1008.2652 [hep-ph]].


\bibitem{Ducati}
  M.~B.~G.~Ducati, M.~T.~Griep and M.~V.~T.~Machado,
  Phys.\ Rev.\ D {\bf 89} (2014) 034022;
  [arXiv:1307.6882 [hep-ph]].


\bibitem{SchSz}
  W.~Sch\"{a}fer and A.~Szczurek,
  Phys.\ Rev.\ D {\bf 93} (2016) 074014;
  [arXiv:1602.06740 [hep-ph]].





\bibitem{Watt:2003vf}
  G.~Watt, A.~D.~Martin and M.~G.~Ryskin,
  Phys.\ Rev.\ D {\bf 70} (2004) 014012;
   Erratum: [Phys.\ Rev.\ D {\bf 70} (2004) 079902];
 [hep-ph/0309096].


\bibitem{HHJ}
  F.~Hautmann, M.~Hentschinski and H.~Jung,
  Nucl.\ Phys.\ B {\bf 865} (2012) 54;
  [arXiv:1205.1759 [hep-ph]].


\bibitem{Dooling:2014kia}
  S.~Dooling, F.~Hautmann and H.~Jung,
  Phys.\ Lett.\ B {\bf 736} (2014) 293;
  [arXiv:1406.2994 [hep-ph]].


\bibitem{Nefedov:2012cq}
  M.~A.~Nefedov, N.~N.~Nikolaev and V.~A.~Saleev,
  Phys.\ Rev.\ D {\bf 87} (2013) 014022;
  [arXiv:1211.5539 [hep-ph]].

\bibitem{Baranov:2014ewa}
  S.~P.~Baranov, A.~V.~Lipatov and N.~P.~Zotov,
  Phys.\ Rev.\ D {\bf 89} (2014) 094025;
  [arXiv:1402.5496 [hep-ph]].




\bibitem{BLZprompt}
  S.~P.~Baranov, A.~V.~Lipatov and N.~P.~Zotov,
  Phys.\ Rev.\ D {\bf 77} (2008) 074024;
 [arXiv:0708.3560 [hep-ph]];
  S.~P.~Baranov, A.~V.~Lipatov and N.~P.~Zotov,
  Eur.\ Phys.\ J.\ C {\bf 56} (2008) 371;
 [arXiv:0805.2650 [hep-ph]].


\bibitem{promptRV}
  S.~Benic, K.~Fukushima, O.~Garcia-Montero and R.~Venugopalan,
  JHEP {\bf 1701} (2017) 115;
 [arXiv:1609.09424 [hep-ph]].



\bibitem{DeakSchwen}
  M.~Deak and F.~Schwennsen,
  JHEP {\bf 0809} (2008) 035;
  [arXiv:0805.3763 [hep-ph]].


\bibitem{BLZdy}
  S.~P.~Baranov, A.~V.~Lipatov and N.~P.~Zotov,
  Phys.\ Rev.\ D {\bf 78} (2008) 014025;
  [arXiv:0805.4821 [hep-ph]].

\bibitem{JAXO}
  D.~Binosi, J.~Collins, C.~Kaufhold and L.~Theussl,
  Comput.\ Phys.\ Commun.\  {\bf 180} (2009) 1709;
  [arXiv:0811.4113 [hep-ph]].


\bibitem{Collins_Soper_frame}
J.~C.~Collins and D.~E.~Soper, Phys.\ Rev.\ D {\bf 16} (1977) 2219.



\bibitem{Faccioli:2011pn}
  P.~Faccioli, C.~Lourenco, J.~Seixas and H.~K.~Wohri,
  Phys.\ Rev.\ D {\bf 83} (2011) 056008;
  [arXiv:1102.3946 [hep-ph]].













\bibitem{Gottfried:1964nx}
  K.~Gottfried and J.~D.~Jackson,
  Nuovo Cim.\  {\bf 33} (1964) 309.



\bibitem{CollEll}
  J.~C.~Collins and R.~K.~Ellis,
  Nucl.\ Phys.\ B {\bf 360} (1991) 3.





\bibitem{EffectiveAction}
  L.~N.~Lipatov,
  Nucl.\ Phys.\ B {\bf 452} (1995) 369;
 [hep-ph/9502308];
  E.~N.~Antonov, L.~N.~Lipatov, E.~A.~Kuraev and I.~O.~Cherednikov,
  Nucl.\ Phys.\ B {\bf 721} (2005) 111;
[hep-ph/0411185];
  L.~N.~Lipatov and M.~I.~Vyazovsky,
  Nucl.\ Phys.\ B {\bf 597} (2001) 399;
 [hep-ph/0009340].


\bibitem{FORM}
  J.~A.~M.~Vermaseren,
  math-ph/0010025.





\bibitem{JHTMD}
  F.~Hautmann and H.~Jung,
  Nucl.\ Phys.\ B {\bf 883} (2014) 1;
[arXiv:1312.7875 [hep-ph]].



\bibitem{TMDlib}
  F.~Hautmann, H.~Jung, M.~Kr\"{a}mer, P.~J.~Mulders, E.~R.~Nocera, T.~C.~Rogers and A.~Signori,
  Eur.\ Phys.\ J.\ C {\bf 74} (2014) 3220;
 [arXiv:1408.3015 [hep-ph]].



\bibitem{MotSadHERA}
  L.~Motyka and M.~Sadzikowski,
  Acta Phys.\ Polon.\ B {\bf 45} (2014) 2079;
  [arXiv:1411.7774 [hep-ph]].


\bibitem{GBW}
  K.~J.~Golec-Biernat and M.~W\"{u}sthoff,
  Phys.\ Rev.\ D {\bf 59} (1998) 014017;
  [hep-ph/9807513];
%
  Phys.\ Rev.\ D {\bf 60} (1999) 114023;
 [hep-ph/9903358].





\bibitem{KwieMaSt}
  J.~Kwieci\'{n}ski, A.~D.~Martin and A.~M.~Sta\'{s}to,
  Phys.\ Rev.\ D {\bf 56} (1997) 3991;
[hep-ph/9703445].




\bibitem{geometric}
  A.~M.~Sta\'{s}to, K.~J.~Golec-Biernat and J.~Kwieci\'{n}ski,
  Phys.\ Rev.\ Lett.\  {\bf 86} (2001) 596;
 [hep-ph/0007192].


\bibitem{KwMoTi}
  N.~Timneanu, J.~Kwieci\'{n}ski and L.~Motyka,
  Eur.\ Phys.\ J.\ C {\bf 23} (2002) 513;
  [hep-ph/0110409].


\bibitem{MMHT} 
  L.~A.~Harland-Lang, A.~D.~Martin, P.~Motylinski and R.~S.~Thorne,
  Eur.\ Phys.\ J.\ C {\bf 75} (2015) 204;
 [arXiv:1412.3989 [hep-ph]].


\bibitem{Kulesza:1999gm}
  A.~Kulesza and W.~J.~Stirling,
  Nucl.\ Phys.\ B {\bf 555} (1999) 279;
 [hep-ph/9902234].









\end{thebibliography}
\end{document}